\documentclass[]{aa}

\usepackage{graphicx}
\usepackage[switch]{linenoaa}
\usepackage{txfonts}

\usepackage[colorlinks=true, linkcolor=blue, citecolor=blue, urlcolor=blue]{hyperref}
\usepackage{orcidlink}

\begin{document}
   \title{High-precision proper motions toward the Galactic bulge Baade's window from a decade of DECam observations}
   \titlerunning{DECam proper motions in the Baade's window}

    \author{
        C. Salvo-Guajardo\inst{1}\thanks{Corresponding author: casalvo@uc.cl}\orcidlink{0009-0007-6076-0115}
        \and
        A. Calamida\inst{2,3}\orcidlink{0000-0002-0882-7702}
        \and
        C.~I. Johnson\inst{2}\orcidlink{0000-0002-8878-3315}
        \and
        M. Dall'Ora\inst{3}\orcidlink{0000-0001-8209-0449}
        \and
        M. Zoccali\inst{1}\orcidlink{0000-0002-5829-2267}
        \and
        A. Luna\inst{4}\orcidlink{0000-0001-5971-8058}
        \and
        M. Gholami\inst{3}
        \and
        J. Olivares Carvajal\inst{3}\orcidlink{0009-0000-1857-4393}
        \and
        T. M. Brown\inst{2}\orcidlink{0000-0002-1793-9968}
        \and
        A. Saha\inst{5}\orcidlink{0000-0002-6839-4881}
        \and  
        M. Catelan\inst{1}\orcidlink{0000-0001-6003-8877}
        \and
        N. Rodríguez-Segovia\inst{6}\orcidlink{0000-0002-0125-1472}
        \and  
        A. Valenzuela Navarro\inst{1}\orcidlink{0000-0002-9523-1228}
        \and
        M. Gennaro\inst{2,7}\orcidlink{0000-0002-5581-2896}
        \and    
        R. Contreras Ramos\inst{1}\orcidlink{0000-0001-7948-9731}
    }

    \institute{
        Instituto de Astrof\'isica, Pontificia Universidad Cat\'olica de Chile,
        Av. Vicu\~na Mackenna 4860, 782-0436 Macul, Santiago, Chile
        \and
        Space Telescope Science Institute - AURA,
        3700 San Martin Drive, Baltimore, MD 21218, USA
        \and
        INAF - Osservatorio Astronomico di Capodimonte,
        salita Moiariello 16, 80131, Naples, Italy
        \and
        European Southern Observatory,
        Alonso de C\'ordova 3107, Vitacura, Santiago, Chile
        \and
        NSF’s National Optical Infrared Astronomy Research Laboratory,
        950 North Cherry Avenue, Tucson, AZ 85719, USA
        \and
        Korea Astronomy and Space Science Institute, 776, Daedeokdae-ro, Yuseong-gu, Daejeon 34055, Republic of Korea
        \and
        The William H. Miller III Department of Physics \& Astronomy,
        Bloomberg Center for Physics and Astronomy, Johns Hopkins University,
        3400 N. Charles Street, Baltimore, MD 21218, USA
    }

   \date{Received 2026, September 19; accepted 2026, November 19}

 
  \abstract
   {A complete characterization of the Galactic bulge stellar populations could offer important clues on the formation and evolution of our Galaxy. However, most bulge fields are hampered by high stellar crowding, strong interstellar extinction, and by heavy contamination from foreground disk stars. Proper motions are the most effective tool to separate the bulge from the disk component and to provide kinematically clean samples of bulge stars. Hubble Space Telescope (HST) has provided pure bulge star samples, reaching well below the main-sequence turnoff of its stellar population, but limited to a few small fields of view.}
   {We aim to derive deep, high-precision proper motions towards the Baade's window by using wide-field ground-based Dark Energy Camera (DECam) observations, extending kinematic studies of the Galactic bulge below the turn-off to a much larger field of view. We also develop a method that could be applied to the study of the Galactic bulge stellar populations with Rubin and Roman surveys in the future.}
   {We combined three datasets observed with DECam in the $r$-band spanning a ${\sim}$~10-year baseline. As a pilot study, we only analyzed the field covered by a single detector. We developed an astrometric calibration pipeline that corrects for the DECam tree-ring pattern and residual distortions, tying all positions to the absolute \textit{Gaia}\,Data Release 3 (DR3) reference frame, and used them to derive proper motions. We isolated the bulge stellar population through a strict cut in the proper motion component along Galactic longitude, refined with a \textit{Gaia}\,DR3 parallax selection.}
    {We measured proper motions for $48\,239$ stars in the pilot field, with a median uncertainty of $0.65$~mas~yr$^{-1}$ at $r_0<18$ mag. Our proper motions agree closely with \textit{Gaia}\,DR3, and have uncertainties smaller than those of the VVV Infrared Astrometric Catalogue 2 (VIRAC2) in the magnitude range $r_0>14$ mag, reaching ${\sim}1.5$ magnitudes deeper than \textit{Gaia} and a comparable depth to VIRAC2 in terms of relative completeness. We found velocity dispersions along Galactic longitude and latitude of $\sigma_{\mu_{\ell^\ast}}=2.76\pm0.03$~mas~yr$^{-1}$ and $\sigma_{\mu_b}=2.49\pm0.02$~mas~yr$^{-1}$ for the Galactic bulge, with an anisotropy of $\sigma_{\mu_{\ell^\ast}}/\sigma_{\mu_b} = 1.11\pm0.02$, consistent with HST studies. We estimated a foreground contamination of ${\sim}14\%$ before any kinematic selection. The cleaned color-magnitude diagram reaches below the main-sequence turnoff, and comparison with isochrones shows no evidence for a large population younger than ${\sim}4$--$5$~Gyr in this field. A small fraction (${\sim}2.5\%$) of stars remains in the foreground blue-sequence region, likely residual disk contaminants and bulge blue stragglers. Moreover, since the Vera C. Rubin Observatory camera shares the same CCD technology as DECam, this work offers a valuable precursor for crowded-field astrometry in the Rubin era.}
    {}
   \keywords{astrometry -- proper motions -- Galaxy: bulge -- Galaxy: kinematics and dynamics}

   \maketitle
   \nolinenumbers

\section{Introduction}

The Galactic bulge serves as an ideal laboratory for studying not only our own Galaxy but also the evolutionary history of galaxies in general, since it is the only bulge we can study in detail, resolving most of its stars individually. It covers approximately the central 2 kpc of the Milky Way (MW) and represents one of its oldest and most massive components, containing about 20\% of its total stellar mass \citep[e.g.,][]{licquia+15,valenti+16,portail+17,simion+17,cautun+20}. Because it contains a significant fraction of the Galaxy's stellar mass and preserves stars formed in early stages, the bulge provides a unique record of the formation of the MW.

Observationally, the Galactic bulge is known to host a bar with a boxy/peanut (B/P) morphology and an X-shaped structure \citep{mcwilliam+10, nataf+10, saito+11, wegg+13, ness+16, clarke+19, sanders+19, marchetti+22, khoperskov+2025}. Such a structure arises naturally from the secular evolution of the disk through bar formation and buckling \citep{raha+91,merritt+94,debattista+06,boin+2024}, or from the resonant trapping of stellar orbits within the bar \citep{combes+81, combes+90, quillen+14, sellwood+20}. This contrasts with the classical bulges of other galaxies, which are thought to form through the hierarchical growth and merging of galaxies \citep{binney+85,kauffmann+93}. However, the MW bulge appears to combine both channels. While its more metal-rich stars trace the B/P structure, a more metal-poor population traces a spheroidal distribution, rotates more slowly, and does not follow the X-shape \citep{dekany+13,kunder+16,zoccali+17,queiroz+2020,lim+2021,olivares+2024,quezada+2025,nepal+2026}, pointing to an older component consistent both with a very weak bar \citep{dimateo+16,debattista+17} or with a classical hierarchically formed bulge \citep{athanassoula+17}. The MW bulge is therefore likely a composite system, shaped by both secular and early processes \citep[e.g.,][]{babusiaux+10,zoccali+14,portail+17}.

However, studying the Galactic bulge involves significant observational challenges. Its intrinsically high stellar density \citep{valenti+16} results in high levels of crowding and source confusion, which require Point-Spread Function (PSF) photometry to accurately resolve individual stars \citep[e.g.,][]{stetson1987}. Additionally, because the Sun is located in the Orion arm at approximately 8 kpc from the Galactic center \citep[e.g.,][]{ghez+08,reid+14,gravity+2022}, observations toward the bulge must traverse the Galactic disk. Consequently, bulge fields experience strong and spatially variable interstellar extinction due to intervening dust and gas \citep{schultheis+99,gonzalez+11,nataf+13,schultheis+14,saha+19}, as well as substantial contamination from foreground disk stars along the line of sight \citep[e.g.,][]{kuijken+02}. These combined effects hinder the ability to obtain clean bulge samples, thereby limiting current understanding of the formation and evolution of the Galactic bulge (see \citealt{barbuy+18,zoccali+26} for recent reviews).

Among the available approaches to mitigate foreground contamination, proper motions (PMs) offer a powerful tool to separate bulge stars from foreground contaminants, such as disk stars, which overlap in color and magnitude but differ in their kinematics. The precision of PM measurements depends on both the astrometric accuracy of individual observations and the temporal sampling, with longer baselines and multiple epochs generally yielding more robust results. This was first acknowledged by \citet{kuijken+02}, who used observations from the Hubble Space Telescope (HST) to kinematically obtain a clean bulge sample, concluding that the old stellar population of the inner bulge is rotating.

The advent of the \textit{Gaia} mission \citep{gaiadr1} revolutionized Galactic astrometry by providing positions, parallaxes, and PMs for over one billion stars in its most recent data release \citep[\textit{Gaia}\,DR3;][]{gaiadr3}. Because \textit{Gaia} operates at optical wavelengths, it is especially susceptible to high, spatially variable reddening along these lines of sight. However, crowding further degrades the quality of astrometric measurements. In the most densely populated bulge fields, \textit{Gaia}\,DR3 PM uncertainties may be underestimated by up to a factor of ${\sim}4$ compared to HST measurements \citep{luna+23}. Furthermore, \textit{Gaia} does not reach the faint bulge main sequence (MS) and only marginally reaches the main sequence turnoff (MSTO), which is critical for deriving robust age constraints. As a result, deep kinematic studies of the most obscured bulge populations remain difficult.

The near-infrared (NIR) regime offers an alternative for studying the bulge, since extinction is roughly an order of magnitude lower than at optical wavelengths. Multi-epoch observations from the Vista Variables in the Vía Láctea survey \citep[VVV;][]{minniti+10} and its extension, VVV eXtended \citep[VVVX;][]{saito+24}, have yielded astrometric surveys of large regions of the bulge. In particular, the VVV Infrared Astrometric Catalogue \citep[VIRAC;][]{smith+18} and its latest release, VIRAC2 \citep{smith2025}, based on approximately 10 years of observations covering more than 500 $\mathrm{deg}^2$ of the Galactic bulge and plane, provide PMs for hundreds of millions of stars in the \textit{Gaia} astrometric system. The VIRAC2 PMs typically achieve uncertainties of ${\sim}0.37$~mas~yr$^{-1}$ for bright stars, but these uncertainties increase to more than 2.5~mas~yr$^{-1}$ for stars near the MSTO \citep{smith2025}.

When studying the Galactic bulge in the optical, it is common to focus on the so-called low-extinction windows—lines of sight with relatively low interstellar extinction—even with optical instruments. Among the most studied fields are the Baade’s window (at $\ell=1.02^\circ$ and $b=-3.92^\circ$) and the Sagittarius window (at $\ell=1.25^\circ$ and $b=-2.65^\circ$). The high spatial resolution of HST has prompted numerous observations in these fields. In fact, there are many works that have used multi-epoch HST observations to derive PMs in these windows, yielding velocity dispersions in the range of $\sigma_{{\mu}_{\ell}^\ast} = 2.5-3.1$~mas~yr$^{-1}$ and $\sigma_{{\mu}_{b}} = 2.4-3.0$~mas~yr$^{-1}$ for the Galactic bulge (e.g., \citealt{kuijken+02,kozlowski+06,clarkson+08,clarkson2011,calamida+2014,terry+20}; see also \citealt{clarke+19} for estimates derived from VIRAC PMs). Furthermore, this has enabled the construction of decontaminated color-magnitude diagrams (CMDs) of the bulge that extend well below the MSTO \citep[e.g.,][]{clarkson+08,clarkson2011,calamida+2014,calamida+2015,bernard+18,renzini+18,terry+20}. These clean CMDs have revealed important features of the bulge population, including a candidate population of blue straggler stars \citep[BSSs;][]{clarkson2011} and the white dwarf cooling sequence \citep{calamida+2014}.

Reaching the MSTO and going below it are key to constraining the bulge star formation history (SFH), since the MSTO is strongly sensitive to stellar age. In fact, many works have consistently supported that the bulge is predominantly old, with ages $\gtrsim 10$~Gyr \citep[e.g.,][]{zoccali2003,clarkson+08,clarkson2011,bernard+18,renzini+18,surot2019}. However, the existence, and magnitude, of a minor younger or intermediate-age component remains debated \citep{haywood+16,bensby+17,joyce+23}. For instance, \citet{clarkson2011} provided an upper limit of ${\sim}3.4\%$ for the bulge population younger than 5 Gyr, while \citet{bensby+17} found that around ${\sim}25\%$ of the metal-rich population is younger than 5 Gyr. Nevertheless, most of these studies are restricted by the small field of view (FoV) of HST \citep{clarkson+08,clarkson2011,calamida+2014,calamida+2015,bernard+18,terry+20}, and a more comprehensive picture of bulge stellar populations would require deep samples over significantly larger sky areas.

\begin{figure*}
    \centering
    \includegraphics[width=\hsize]{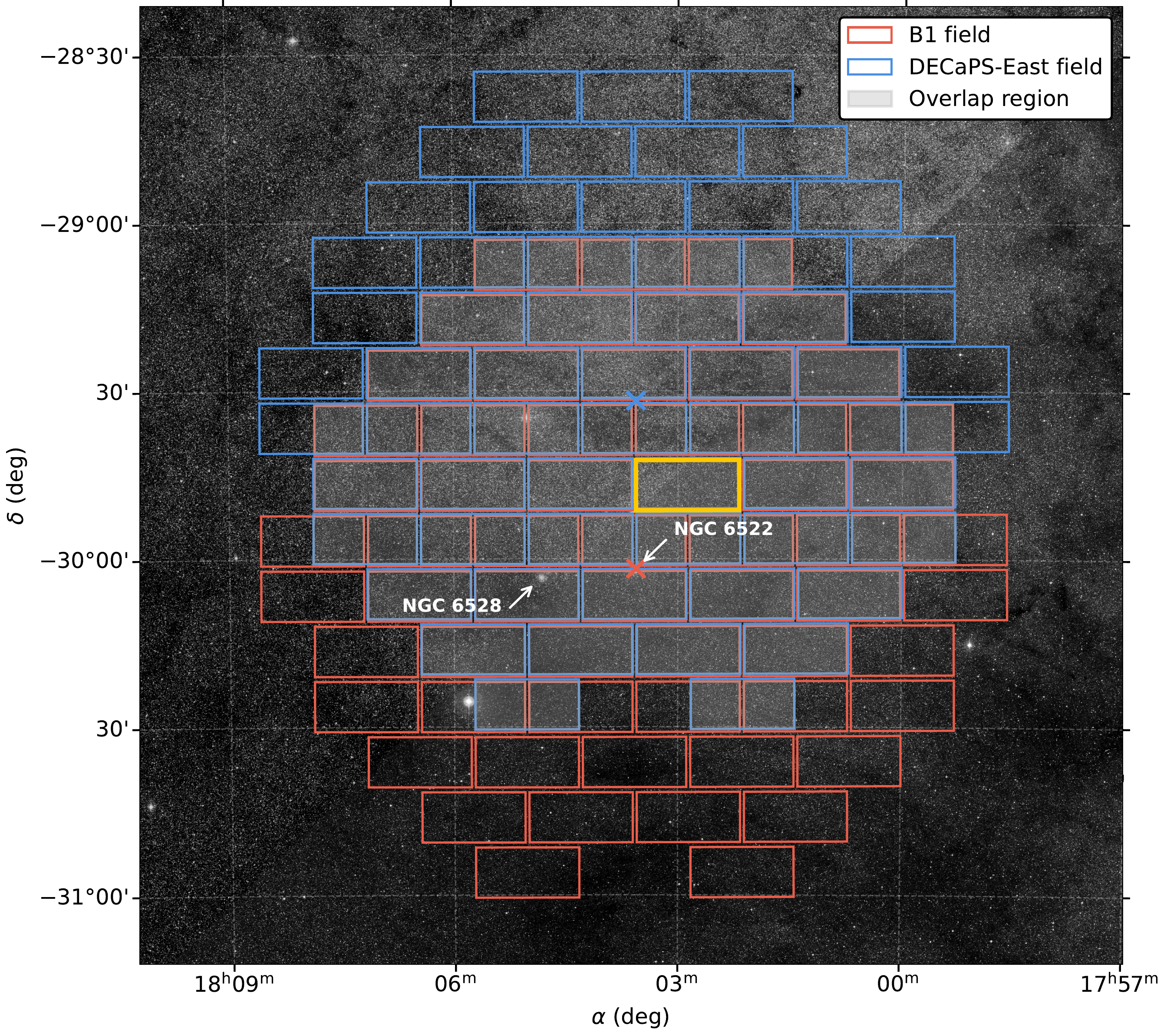}
    \caption{Sky coverage of the B1 field (red) and the DECaPS-East field (blue) in the Baade's window. The grey shaded fields indicate the overlapping region between both fields. The red cross at $(\alpha, \delta) = (270.89^\circ, -30.03^\circ)$ marks the center of the FoV of B1, while the blue cross at $(\alpha, \delta) = (270.89^\circ, -29.53^\circ)$ represents the center of the FoV of DECaPS-East. The yellow-highlighted rectangle represents the pilot field analyzed in this work (detector S10 in B1 and N10 in DECaPS-East). White arrows mark the positions of the two globular clusters present in this field, NGC 6522 and NGC 6528. In the background there is a sky image taken from the Digital Sky Survey in the r-band.}
    \label{Fig:overlap_region}
\end{figure*}

Such wide-area coverage is, at the moment, only feasible from the ground. In the NIR regime, VIRAC2 provides PMs over much of the Galactic bulge, but it has limited precision near the MSTO. At optical wavelengths, the Dark Energy Camera \citep[DECam;][]{decam2008,decam2015} on the Blanco 4-m telescope at CTIO offers a complementary route, combining a wide FoV (${\sim}3$~deg$^2$) with a small pixel scale (${\sim}0.26 \arcsec \mathrm{px^{-1}}$) and good image quality. These capabilities have already enabled several photometric studies of the Galactic bulge. For example, \citet{saha+19} obtained deep optical photometry for six bulge fields, including the Baade's window, reaching the MSTO and enabling detailed studies of bulge stellar populations. Subsequently, the Blanco DECam Bulge Survey \citep[BDBS;][]{rich+2020,johnson+2020} extended DECam bulge observations to more than ${\sim}200$~deg$^2$ across the southern bulge. More recently, \citet{marchetti+22} combined BDBS photometry with \textit{Gaia}\,EDR3 astrometry to obtain clean bulge samples over a large fraction of the survey area, finding little evidence for a widespread young bulge population and reinforcing the picture of a predominantly old bulge. However, since the analysis relied on \textit{Gaia} astrometry, it was limited by its depth and completeness in the Galactic bulge.

Despite these advances, no deep PM catalog has yet been derived directly from DECam observations. Constructing such a catalog requires careful control of the instrumental systematics affecting DECam astrometry, particularly the optical field distortion and the detector-level \textit{tree-ring} pattern, a source of radial astrometric shifts caused by doping variations in thick, fully depleted CCDs \citep{plazas2014,bernstein2017}. This effort is particularly relevant in the era of the Legacy Survey of Space and Time \citep[LSST;][]{ivezic+2019}, whose camera is based on the same CCD technology and thus is affected by similar systematics \citep{esteves2023}. Therefore, DECam also serves as an important precursor for the development and validation of crowded-field astrometric techniques for future LSST studies of the Galactic bulge.

In this work, we present high-precision PMs for stars in the Baade's window built from a decade of DECam observations. We develop a pipeline that corrects systematic astrometric effects and computes PMs directly tied to the \textit{Gaia}\,DR3 absolute frame. Here, we restrict our analysis to one field covered by a single DECam detector. The resulting PMs reach at least one magnitude below the \textit{Gaia} limit in this field and attain uncertainties that are competitive with, and on average smaller than, those of VIRAC2 in the MSTO, enabling a kinematically clean CMD of the bulge population. The paper is organized as follows: Sect.~\ref{sec:data} describes the observations and the photometry. Sect.~\ref{sec:astrometric_calibration} details the astrometric calibration, including the correction of systematics. Sect.~\ref{sec:pms} derives the PMs. Sect.~\ref{sec:gaia} presents comparisons with \textit{Gaia}\,DR3 and VIRAC2. Sect.~\ref{sec:kinematics} analyzes the Galactic bulge and foreground disk stellar kinematics. Sect.~\ref{sec:bulge_decontamination} presents a kinematic decontamination for Galactic bulge stars. Finally, Sect.~\ref{sec:summary} summarizes our work.

\section{Data} \label{sec:data}

    \begin{table*}[ht]
        \caption{Summary of the data used in this work.}
        \label{Tab:data}
        \centering
        \begin{tabular}{c l l l l c c c}
            \hline\hline\noalign{\smallskip}
            Epoch & Dataset & Date range & PI & Program ID & Filter & Exposure time (s) & Number of images \\
            \noalign{\smallskip}\hline\noalign{\smallskip}
            2013 & B1               & May--Jun 2013 & Saha   & 2013A-0719   & r & 100 & 29  \\
            2021 & DECaPS-East 2021 & Mar--May 2021 & Graham & 2021A-0113   & r & 50  & 175 \\
            2023 & DECaPS-East 2023 & Mar--Aug 2023 & Nir    & 2023A-716082 & r & 50  & 202 \\
            \hline
        \end{tabular}
        \tablefoot{The table lists the observing epochs, dataset labels, date ranges, principal investigators, program IDs, filters, exposure times, and number of images included in the analysis.}
    \end{table*}

To calculate the PM of an object we need to know its position in at least two different epochs. In this study, we used three epochs of observations spanning a 10 year baseline, drawn from two independent observing programs. The first epoch belongs to the program presented in \citet{saha+19}, while the second and third epochs were collected as part of the Deep Drilling Field (DDF) program of the DECam Alliance for Transients \citep[DECAT;][]{graham+23,graham+2024}.

The three datasets are centered on the low-reddening Baade's window and were collected with DECam in the \textit{ugriz} filters. DECam is a wide field optical camera mounted on the Blanco 4-m Telescope located at Cerro Tololo Inter-American Observatory (CTIO) in Chile and operated by NOIRLab (AURA). It has a FoV of 132$' \times$ 132$'$ with a pixel size of $0.2637\arcsec\mathrm{px}^{-1}$ at the center and $0.2626\arcsec\mathrm{px}^{-1}$ at the edges, and a total of 62 CCDs, one of which has been non-functional since 2012, while another one is affected by amplifier issues\footnote{\url{https://noirlab.edu/science/programs/ctio/instruments/Dark-Energy-Camera/Known-Problems}}. Neither of them corresponds to the CCD analyzed in this work. The detectors are composed of 2048$\times$4096 red-sensitive pixels with an inter-CCD gap of 201 pixels along the long edge (i.e., the horizontal inter-CCD gaps in Fig.~\ref{Fig:overlap_region}) and $153$ pixels along the short edge (i.e., the vertical inter-CCD gaps in Fig.~\ref{Fig:overlap_region}). There are no overlaps between the different CCDs from the same dataset.

\subsection{First epoch dataset}

The first epoch corresponds to observations carried out in 2013 (program ID 2013A-0719), which constituted the basis of the work by \citet{saha+19}, where this field is labeled as B1. Although these observations were obtained in all the \textit{ugriz} filters, for the PM measurements we only consider the long-exposure images (100\,s), taken in the $r$-band, which provide the best signal-to-noise ratio in this dataset. This is essential to reach the faintest possible magnitudes and extend our PMs below the MSTO. These data comprise 29 images observed between May and June, 2013. The FoV of these observations is centered on the Baade's window, at $(\alpha , \delta) = (270.89^{\circ}, -30.03^{\circ})$. Fig.~\ref{Fig:overlap_region} shows the FoV covered by these observations (fields marked in red), with the center indicated by a red cross.

\subsection{Second epoch dataset}

The second epoch corresponds to observations of the DECaPS-East field from the DDF program \citep[][Rodríguez-Segovia et al., \textit{in prep.}]{graham+23}, obtained in 2021 (program ID 2021A-0113; hereafter ``DECaPS-East 2021''). As in the first epoch, this dataset covers multiple filters (\textit{griz}), but we restrict our analysis to the longest-exposure (50\,s) frames in the $r$ band, amounting to 175 images acquired between March and May, 2021. This FoV is centered at $(\alpha , \delta) = (270.89^{\circ}, -29.53^{\circ})$ and only partially overlaps with the first epoch, with a declination offset of about $+0.50^{\circ}$. The corresponding footprint is also displayed in Fig.~\ref{Fig:overlap_region} (fields marked in blue), with its center indicated by a blue cross.

\subsection{Third epoch dataset}

The third epoch corresponds to observations of the same DECaPS-East field carried out in 2023 (program ID 2023A-716082; hereafter ``DECaPS-East 2023''), which is also included in the work of \citet{graham+23} and Rodríguez-Segovia et al. (\textit{in prep.}). As for the second epoch, these data span the \textit{griz} filters, but only the longest-exposure (50\,s) images in the $r$ band are considered here, for a total of 202 frames taken between March and August, 2023. This dataset covers exactly the same field as the second epoch, so its footprint is also shown in blue in Fig.~\ref{Fig:overlap_region}, with the center indicated by a blue cross.

\subsection{Data reduction}

Rather than processing the full DECam footprint, in this work we focus on a single CCD (marked in yellow in Fig.~\ref{Fig:overlap_region}), which we adopt as a pilot field to test our methodology. Additionally, we note that no dithering pattern was applied. Within each night, the pointings differ by only a few arcseconds (median dispersion of ${\sim}12\arcsec$ in 2013, and ${\sim}3.6\arcsec$ in 2021 and 2023).

The images from the three epochs were processed through the DECam community pipeline \citep[DCP;][]{valdes+2014}, which performs the instrumental signature removal and the initial astrometric calibration. The photometry was then performed on these images in the context of the GalActic bulGe with plEiadi (GAGE) project \citep{gholami+26}, a time-series exploration of the Galactic bulge based on archive DECam data. The details of the photometry and of the calibration are discussed in \citet{gholami+26}, and in Luna et al. (\textit{in prep.}). Briefly, all individual images were reduced using \textsc{DAOPHOT/ALLSTAR}, which carries out PSF photometry suited to crowded stellar fields \citep{stetson1987,stetson1994}. The PSF was constructed through an iterative procedure that rejected outlying PSF stars, with the analytical fitting function selected to minimize the fit residuals. We adopted a spatially constant PSF, as allowing linear or quadratic spatial variations did not significantly improve the photometry. Seeing conditions were comparable across the three datasets, with an overall median FWHM of $3.87$~px ($1.05\arcsec$). In principle, both seeing and crowding can have an impact on the photometric centroids of the sources. However, we expect that this can affect the dispersion of the centroid measurements without introducing systematics. Finally, the sources detected in the different images were cross-matched with the routines \textsc{DAOMATCH/DAOMASTER} \citep{stetson93}, assigning unique and consistent IDs across epochs so that each star can be tracked from one epoch to the next.

From these catalogs, we selected sources with $\chi^2 <5$ and $|\mathrm{sharpness}|<1$. These thresholds are deliberately permissive, since stricter cuts would preferentially remove the faintest sources, which are those we aim to reach below the MSTO. In particular, the sharpness criterion removes non-stellar detections and obvious blends, while residual poorly fitted sources contribute with correspondingly lower weight in the subsequent analysis, since both the astrometric fits and the PM determination are error-weighted and include iterative $3\sigma$ clipping. Note that these catalogs are used exclusively for the astrometric analysis, while the photometry included in all the CMDs comes from the extinction-corrected catalog of \citet{saha+19} (see Sect.~\ref{sec:pms}).

A summary of the different datasets used in this work is presented in Table~\ref{Tab:data}. For further details on the observations, we refer the reader to \citet{saha+19} for B1 and \citet{graham+23} for the DECaPS-East epochs, as appropriate.

\section{Astrometric calibration}
\label{sec:astrometric_calibration}

Our analysis is based on the individual resampled images from the DECam Community Pipeline \citep[DCP;][]{valdes+2014}. The DCP performs instrumental signature removal---including bias, crosstalk, linearity, flat-field, and fringe corrections---followed by an astrometric calibration in which source catalogs extracted by \textsc{SExtractor} \citep{bertin+1996} are matched to a reference catalog using \textsc{SCAMP} \citep{bertin2006}, yielding a world coordinate system (WCS) that includes a polynomial distortion model combined with a tangent-plane projection.

The three observational epochs considered in this work were processed with different DCP versions and astrometric reference catalogs. The B1 dataset was reduced with PLVER~V4.8.1, in which the astrometric calibration is tied to \textit{Gaia}~DR1 \citep{gaiadr1}. In contrast, the DECaPS-East datasets were processed with more recent pipeline versions (PLVER~5.2 and 5.5, respectively), for which the astrometric solution is tied to \textit{Gaia}\,DR3 \citep{gaiadr3}.

The resampled data products used in this work are generated for each individual exposure by reprojecting each CCD to a standard orientation (north up, east left) at a uniform pixel scale of $0.27\arcsec\mathrm{px}^{-1}$. This reprojection uses the full WCS solution, incorporating the nominal large-scale optical distortion of the DECam wide-field corrector and yielding images whose WCS is approximately described by a linear tangent-plane projection. We note that no coadded images combining multiple exposures were used in our analysis.

However, two important systematics remain in these resampled images. First, the DCP astrometric solution exhibits a systematic offset of ${\sim}0.2$--$0.4\arcsec$ with respect to the \textit{Gaia}\,DR3 reference frame. These offsets are present across all datasets, including those calibrated against \textit{Gaia}\,DR3, indicating that they do not arise from the choice of astrometric reference catalog, but rather from limitations of the global WCS model used by the pipeline. Second, the resampling preserves the tree-ring pattern, a radial astrometric systematic intrinsic to thick, fully depleted CCDs \citep{plazas2014, bernstein2017}, although its amplitude and shape may be partially altered by the resampling process. We describe and model this systematic effect in Sect.~\ref{subsec:treerings}.

To address these systematics, we developed a three-stage calibration pipeline: (i)~a preliminary linear alignment to the \textit{Gaia}\,DR3 reference frame (Sect.~\ref{subsec:preliminary}), (ii)~a correction for the tree-ring pattern (Sect.~\ref{subsec:treerings}), and (iii)~a polynomial transformation from tree-ring-corrected coordinates to the \textit{Gaia} reference frame (Sect.~\ref{subsec:chebyshev}). The associated astrometric uncertainties are described in Sect.~\ref{subsec:uncertainties}.

\subsection{Preliminary alignment to \textit{Gaia}\,DR3}
\label{subsec:preliminary}

To establish an initial alignment between our image positions and the \textit{Gaia}\,DR3 reference frame, we first constructed \textit{Gaia} reference catalogs at the epoch of each observation. We selected sources with $RUWE < 1.4$, as recommended by \citet{lindegren+2021}, \citet{fabricius+2021}, and \citet{stassun+2021}. We additionally applied a brightness cut of $G > 15$ to exclude the brightest sources, which are saturated in our long-exposure DECam images. Since the \textit{Gaia} reference epoch is 2016.0, we propagated all positions to the epoch of each observation using \textit{Gaia} PMs.

The \textit{Gaia} reference positions were transformed from equatorial coordinates $(\alpha, \delta)$ to resampled image pixel coordinates $(X_{\mathrm{\textit{Gaia}}}, Y_{\mathrm{\textit{Gaia}}})$ using the inverse DCP WCS solution and cross-matched with our DECam catalog using a $1\arcsec$ matching radius. To ensure reliable matches, we retained only sources with photometric uncertainties below $0.05$~mag in both datasets and color differences $|G - r| < (G - r)_\mathrm{peak} \pm 0.5$~mag, where $(G - r)_\mathrm{peak}$ is the modal color. This criterion exploits the similarity between the central wavelengths of the DECam $r$-band and the \textit{Gaia} $G$-band \citep{george+2020}.

The preliminary alignment was modeled as a six-parameter linear transformation with iterative $3\sigma$ clipping:

\begin{align}
X_{\mathrm{lin}} &= a_0 + a_1 X + a_2 Y, \\
Y_{\mathrm{lin}} &= b_0 + b_1 X + b_2 Y,
\end{align}

\noindent mapping the image pixel positions $(X, Y)$ onto the \textit{Gaia}-predicted pixel positions $(X_{\mathrm{\textit{Gaia}}}, Y_{\mathrm{\textit{Gaia}}})$. This transformation refines the pointing, rotation, and plate scale of the image positions relative to the more precise \textit{Gaia}\,DR3 reference frame, while its limited flexibility (linear transformation) prevents it from absorbing the high-frequency radial oscillations characteristic of the tree-ring pattern. The pixel-frame residuals from this fit,

\begin{equation}
\Delta X = X_{\mathrm{lin}} - X_{\mathrm{\textit{Gaia}}}, \qquad
\Delta Y = Y_{\mathrm{lin}} - Y_{\mathrm{\textit{Gaia}}},
\end{equation}

\noindent were pooled across all epochs within each dataset to build the tree-ring model described in Sect.~\ref{subsec:treerings}.

\subsection{Tree-ring correction}
\label{subsec:treerings}

\begin{figure}
    \centering
    \includegraphics[width=\hsize]{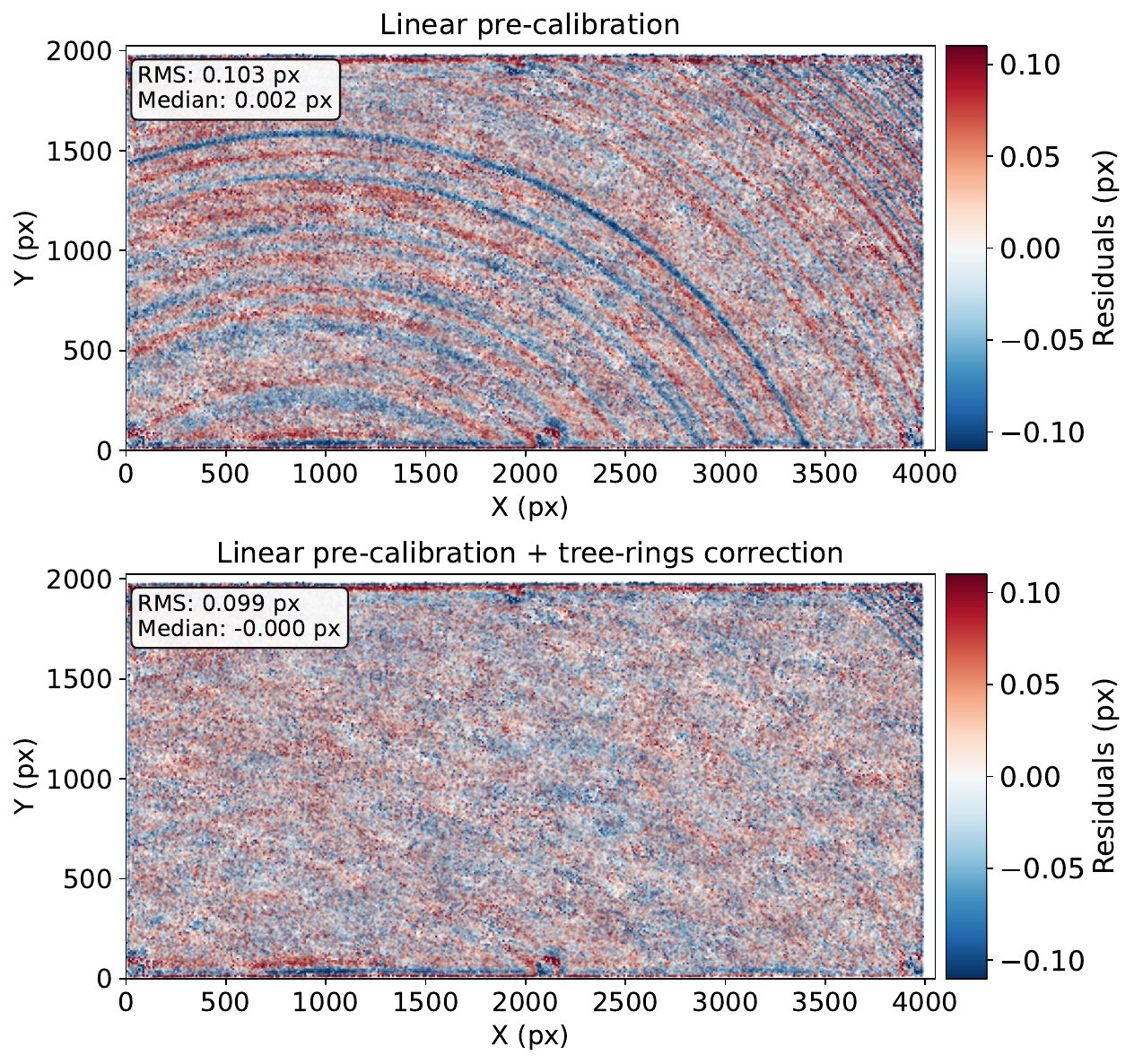}
    \caption{Astrometric residual maps for detector S10 constructed by combining astrometric residuals from 29 individual exposures of the B1 dataset. The color scale shows the radial component  of the position residual with respect to \textit{Gaia}\,DR3. \textit{Top row:} astrometric residuals after the preliminary linear alignment. \textit{Bottom row:} residuals after applying the radial tree-rings correction.}
    \label{fig:tr_residuals_s10}
\end{figure}

\begin{figure*}
    \centering
    \includegraphics[width=\hsize]{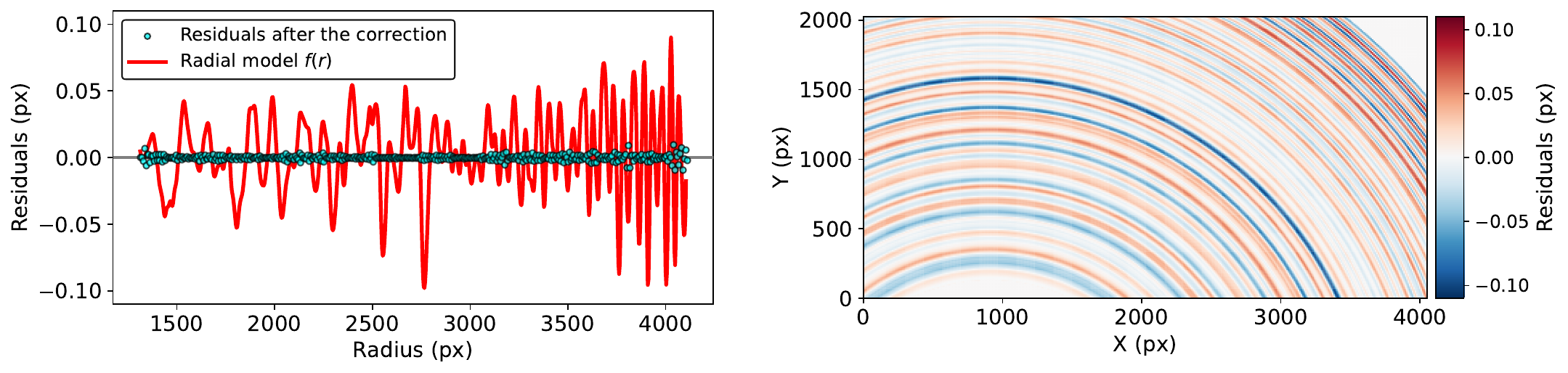}
    \caption{Tree-rings model for detector S10 constructed by combining astrometric residuals from 29 individual exposures of the B1 dataset. \textit{Left panel:} radial model $f(r)$ obtained from a spline fit to the radial residuals, shown in red. Radial astrometric residuals after the tree-rings correction are shown as cyan points. \textit{Right panel:} spatial map of the tree-rings model, color-coded by the model amplitude.}
    \label{fig:tr_model_s10}
\end{figure*}

Although the linear transformation successfully refines the geometric alignment, the stacked pixel-frame residuals reveal systematic patterns that cannot be captured by any smooth global transformation. As shown in the top row of Fig.~\ref{fig:tr_residuals_s10}, the residuals exhibit a clear concentric arc pattern with amplitudes of ${\sim}0.04$--$0.11$~px (${\sim}10$--$30$~mas). This structure corresponds to the so-called tree rings, which arise from radial variations in the dopant concentration during silicon wafer growth \citep[e.g.,][]{plazas2014, bernstein2017}. These induce transverse electric fields that displace the collected charge before it reaches the pixel wells, so the effect is not fully removed by the geometric reprojection applied during the DCP resampling step.

Since the pattern is a fixed property of the silicon crystal, the induced displacement is intrinsically radial in the original detector pixel frame: the position of each star is shifted along the direction connecting its position to the center of the wafer growth pattern. Following \citet{bernstein2017}, we exploit this physical constraint by modeling the correction as a purely radial vector field, but in the resampled image plane:

\begin{equation}
\vec{d}(X, Y) = f(r)\,\hat{r},
\label{eq:tr_radial}
\end{equation}

\noindent where $r = \sqrt{(X - c_X)^2 + (Y - c_Y)^2}$ is the radial distance from the ring center $(c_X, c_Y)$ in pixel coordinates, $\hat{r}$ is the radial unit vector, and $f(r)$ is a scalar function of radius. Projecting onto the pixel axes, the correction can be written as

\begin{align}
\Delta X_{\mathrm{TR}} &= f(r)\,\frac{X - c_X}{r}, \label{eq:tr_dx} \\
\Delta Y_{\mathrm{TR}} &= f(r)\,\frac{Y - c_Y}{r}. \label{eq:tr_dy}
\end{align}

The center of the ring $(c_X, c_Y)$ was determined by maximizing the variance of the azimuthally averaged radial profile. For a candidate center, we computed the radial projection $\Delta r = \Delta X \cos\theta + \Delta Y \sin\theta$, binned it as a function of $r$, and evaluated the variance of the resulting median profile. A high variance indicates that the binned profile coherently traces the ring oscillations, which occurs only near the true center.

For detector S10, the derived center is $(c_X, c_Y) \approx (924, -1099)$~px, located well outside the detector boundary and consistent with the open-arc morphology seen in the top panel of Fig.~\ref{fig:tr_residuals_s10}. For detector N10, the center is $(c_X, c_Y) \approx (3067, 1021)$~px, located inside the FoV (see Appendix~\ref{App:N10}). Since the ring center is a fixed physical property of the silicon wafer, the same ring center is recovered for the N10 detector in both DECaPS-East datasets, as expected.

The radial function $f(r)$ was determined in two steps. First, the radial projection $\Delta r$ was binned into radial intervals, and the median was computed in each bin, requiring a minimum of 15 stars per bin. Second, a cubic smoothing spline was fitted to the binned medians using \textsc{LSQUnivariateSpline} from \textsc{SciPy}, weighted by their number of stars. The resulting spline defines the radial correction function $f(r)$ shown in the left panel of Fig.~\ref{fig:tr_model_s10}, which is evaluated at the radius of each source and applied as a purely radial displacement relative to the fitted tree-ring center. 

The right panel of Fig.~\ref{fig:tr_model_s10} shows the corresponding two-dimensional detector map obtained by evaluating $f(r)$ over the resampled CCD image grid. The alternating positive and negative bands, together with their nearly concentric geometry, are the expected signature of tree-ring distortions in thick CCDs and are consistent with the pattern observed in the stacked residual maps. Here, the upper-right corner of the CCD was excluded from the model due to insufficient angular coverage at larger radii.

The tree-ring-corrected pixel coordinates are therefore

\begin{align}
X_{\mathrm{corr}} &= X - \Delta X_\mathrm{TR}, \label{eq:tr_corr_x} \\
Y_{\mathrm{corr}} &= Y - \Delta Y_\mathrm{TR}. \label{eq:tr_corr_y}
\end{align}

The effect of this correction is illustrated in Fig.~\ref{fig:tr_residuals_s10}. After applying the radial model, the bottom panel shows that the concentric arc pattern is strongly suppressed relative to the preliminary linear solution in the top panel. This behavior is consistent with the left panel of Fig.~\ref{fig:tr_model_s10}, where the post-correction residuals cluster around zero once the fitted radial component is removed.

We acknowledge that correcting the tree-ring pattern in the resampled plane is not ideal, since small misalignments between individual images could partially wash out the signal. Nevertheless, the clear concentric patterns observed in Fig.~\ref{fig:tr_residuals_s10} indicate that the tree-ring signal remains detectable and can be modeled. However, our approach may not fully recover the original detector-level distortion.

\subsection{Polynomial calibration to the \textit{Gaia} reference frame}
\label{subsec:chebyshev}

\begin{figure*}
    \centering
    \includegraphics[width=\hsize]{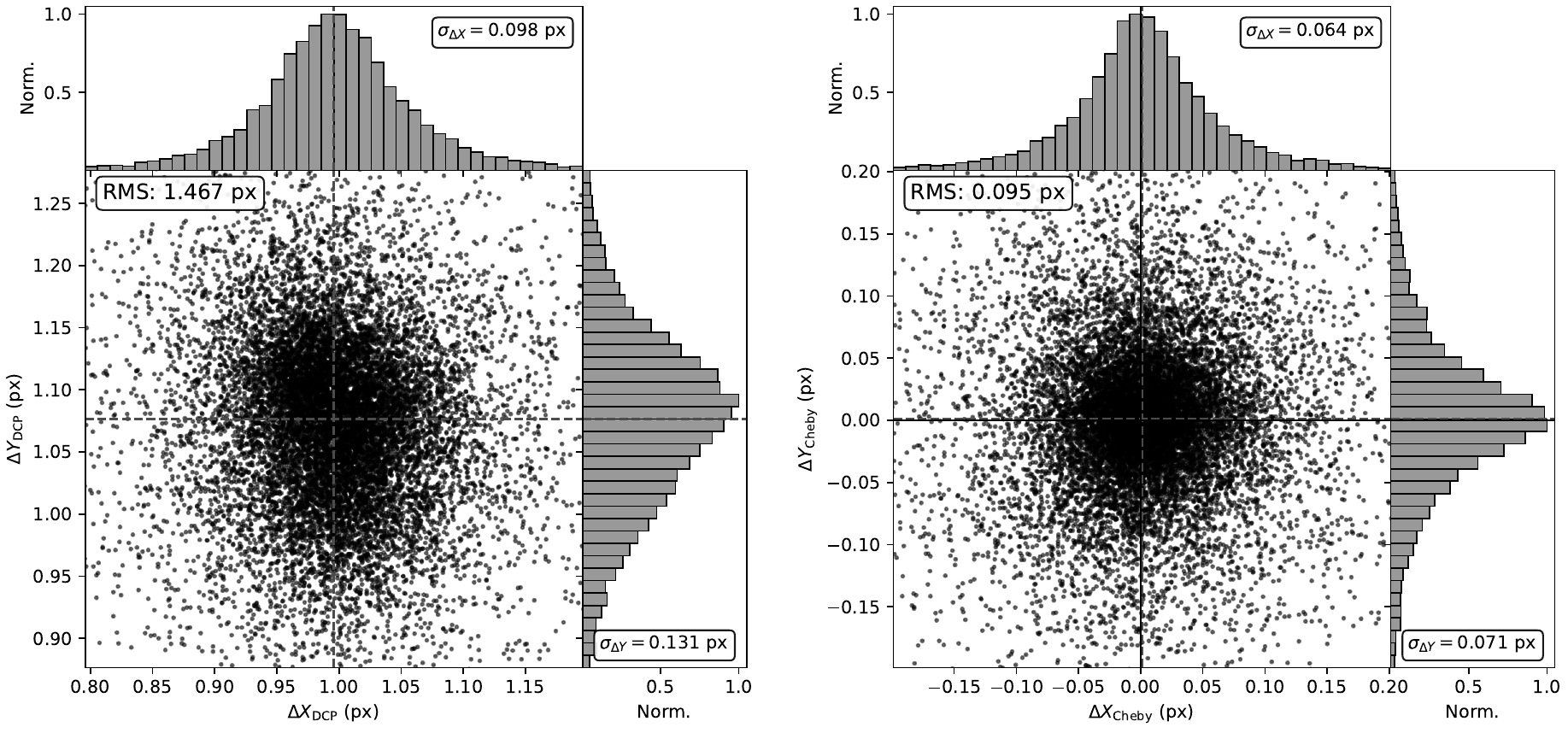}
    \caption{Astrometric residuals for detector S10 from the first observation of 2013 with respect to \textit{Gaia}\,DR3. In both panels, tree-ring corrections have been applied to the data. The left panel shows residuals using the original DCP resampled positions, while the right panel shows residuals after applying additional 4th-order Chebyshev polynomial transformations on top of the tree-ring-corrected positions. Marginal histograms display normalized distributions with their corresponding standard deviations, and dashed lines indicate median offsets.}
    \label{Fig:chebyshev_calibration}
\end{figure*}

After correcting for the tree-ring pattern in the resampled image plane, we transformed the corrected positions $(X_{\mathrm{corr}}, Y_{\mathrm{corr}})$ to equatorial coordinates in the \textit{Gaia}\,DR3 system. Rather than relying on the DCP WCS, we followed the approach of \citet{smith2025} for VIRAC2 and performed this transformation using Chebyshev polynomials in the tangent plane via a gnomonic projection, thereby replacing the DCP WCS with a direct calibration to the \textit{Gaia}\,DR3 frame.

The \textit{Gaia} reference coordinates $(\alpha, \delta)$ are projected onto the tangent plane centered at the geometric center of each CCD $(\alpha_0, \delta_0)$, yielding standard coordinates $(\xi, \eta)$.

The transformation from tree-ring-corrected pixel coordinates to tangent-plane coordinates $(\xi_{\mathrm{cal}}, \eta_{\mathrm{cal}})$ is modeled using two-dimensional Chebyshev polynomials:

\begin{equation}
\tilde\xi_{\mathrm{cal}} =
    \sum_{i+j \leq n} a_{ij}\,T_i(\tilde{X}_{\mathrm{corr}})\,T_j(\tilde{Y}_{\mathrm{corr}}),
\end{equation}

\begin{equation}
\tilde\eta_{\mathrm{cal}} =
    \sum_{i+j \leq n} b_{ij}\,T_i(\tilde{X}_{\mathrm{corr}})\,T_j(\tilde{Y}_{\mathrm{corr}}),
\end{equation}

\noindent where $T_i$ denotes the Chebyshev polynomial of degree $i$, and the tilde indicates that the corresponding coordinates have been normalized to the range $[-1, 1]$ prior to fitting. We adopted $n = 4$, as higher orders did not improve the solution and risked overfitting.

Because the tree-ring signal has been largely corrected in the resampled image pixel coordinates, the polynomial transformation only needs to capture the remaining smooth, low-frequency distortions between the DCP astrometric solution and the \textit{Gaia}\,DR3 reference frame, which are well described by low-order terms. Note that applying the transformation only after correcting the tree-ring pattern also minimizes the possibility of partially absorbing the radial signal into the smooth distortion model. Finally, the calibrated tangent-plane coordinates are denormalized and transformed back to equatorial coordinates.

Polynomial coefficients were fitted via weighted least-squares minimization, using the \textit{Gaia}\,DR3 astrometric uncertainties as weights, with iterative $3\sigma$ clipping to reject outliers, typically converging after 5--7 iterations. The combined calibration (tree-ring correction plus Chebyshev transformation) reduced astrometric residuals from ${\sim}0.20$--$0.40\arcsec$ (${\sim}0.75$--$1.50$~px) to ${\sim}0.02$--$0.04\arcsec$ (${\sim}0.08$--$0.15$~px), achieving $>90\%$ RMS improvement across nearly all observations.

Fig.~\ref{Fig:chebyshev_calibration} illustrates this calibration for detector S10 from the first observation of the B1 field. The RMS decreases from $1.467$~px to $0.095$~px, and the standard deviation drops from ${\sim}0.098$~px in $X$ and ${\sim}0.131$~px in $Y$ before calibration to ${\sim}0.064$~px in $X$ and ${\sim}0.071$~px, respectively. The residual distribution transitions from an irregular, asymmetric shape to a well-concentrated Gaussian distribution centered near zero, with a median offset consistent with zero. This indicates that the dominant residual systematics have been effectively removed.

\subsection{Astrometric uncertainty model}
\label{subsec:uncertainties}

For each epoch, we modeled the total astrometric uncertainty in each coordinate as the quadrature sum of a statistical centroiding uncertainty and a systematic noise floor. Since \textsc{DAOPHOT} does not provide positional uncertainties directly, we estimated the statistical astrometric uncertainty, $\sigma_\mathrm{ast}$, from the instrumental magnitude uncertainty.

For an approximately Gaussian PSF, the noise-limited centroiding precision scales as $\sigma_{\mathrm{ast}} \simeq \sigma_\mathrm{PSF}/\mathrm{SNR}$, where $\sigma_\mathrm{PSF}=\mathrm{FWHM}/2.355$. In addition, from error propagation applied to the magnitude-flux relation, the uncertainty of a magnitude measurement can be expressed in terms of the signal-to-noise ratio as $\sigma_\mathrm{phot}\simeq1.086/\mathrm{SNR}$. Combining these expressions yields:

\begin{equation}
    \sigma_{\mathrm{ast}} \approx 0.39 \times \mathrm{FWHM} \times \sigma_{\mathrm{phot}},
\end{equation}

\noindent where $\sigma_{\mathrm{phot}}$ is the \textsc{DAOPHOT} instrumental magnitude uncertainty and FWHM is the full width at half maximum of the PSF in arcseconds. This relation follows the centroiding formalism of \citet{lindegren1978}, who explicitly considered a two-dimensional Gaussian profile as representative of faint seeing-limited stellar images. In our case, it provides a first-order estimate of the statistical centroiding uncertainty appropriate for the seeing-limited DECam images. Deviations from the ideal Gaussian, noise-limited case, including residual PSF-modeling and other astrometric effects, are accounted for empirically below.

We therefore modeled the total positional uncertainty by adding a systematic noise floor, $\sigma_\mathrm{floor}$, and allowing a multiplicative scaling ($\lambda$) of the statistical variance:

\begin{equation}
    \sigma_{\alpha^\ast} = \sqrt{\sigma_{\mathrm{floor},\alpha^\ast}^2 + \lambda_{\alpha^\ast}\,\sigma_{\mathrm{ast}}^2},
    \label{eq:uncertainty_model}
\end{equation}

\begin{equation}
    \sigma_{\delta} = \sqrt{\sigma_{\mathrm{floor},\delta}^2 + \lambda_{\delta}\,\sigma_{\mathrm{ast}}^2}.
    \label{eq:uncertainty_model_2}
\end{equation}

The parameters $\sigma_\mathrm{floor}$ and $\lambda$ were fitted independently for each coordinate. Specifically, following the same methodology as \citet{smith2025}, we compared our final calibrated stellar positions with the corresponding \textit{Gaia}\,DR3 positions propagated to each observation epoch. The positional residuals were normalized by the quadrature sum of the \textit{Gaia}\,DR3 positional uncertainties and our modeled astrometric uncertainties. We then optimized $\sigma_\mathrm{floor}$ and $\lambda$ in equal-width magnitude bins, requiring the normalized residuals to have a dispersion close to unity, estimated using the median absolute deviation as a robust estimator of the dispersion.

The model was calibrated using the final residuals after applying both the tree-ring and Chebyshev corrections, so that $\sigma_{\mathrm{floor}}$ reflects the systematic floor of the fully calibrated astrometry.

\section{Proper motions}
\label{sec:pms}

PMs were determined from long-exposure $r$-band observations spanning temporal baselines of up to 10~years. While \textit{Gaia}\,DR3 provides exquisite astrometry and PMs for brighter stars, it does not reach below the MSTO in the inner bulge region. We therefore used long-exposure observations to extend our PM measurements to fainter magnitudes.

Given the high cadence within each epoch (see Table~\ref{Tab:data}), the data were binned by month to reduce computational cost and mitigate the impact of correlated systematics. For each monthly bin, we computed the error-weighted mean $\alpha^\ast$ and $\delta$ positions and the corresponding error-weighted mean observation time, after applying $3\sigma$ clipping to reject outliers. The uncertainty on each binned position was obtained by propagating the per-star uncertainties $(\sigma_{\alpha^\ast}, \sigma_{\delta})$ defined in Eqs.~(\ref{eq:uncertainty_model}) and (\ref{eq:uncertainty_model_2}), yielding a combined uncertainty $\sigma_{\mathrm{bin}}$ for each coordinate. The time span of each bin is sufficiently short that stellar motion within a bin has a negligible effect on the astrometry. A characteristic bulge PM dispersion of $2.5$~mas~yr$^{-1}$ corresponds to only ${\sim}0.21$~mas over one month, or ${\sim}8\times10^{-4}$~px at the DECam pixel scale, much smaller than the positional uncertainties of individual measurements.

Then, PMs were derived from linear fits to the monthly-averaged positions as a function of time, for all stars with at least three valid measurements spanning a temporal baseline of at least 8~years. Since all positions were previously calibrated to the \textit{Gaia}\,DR3 reference frame, the resulting PMs are directly tied to the absolute \textit{Gaia}\,DR3 astrometric system. The fitting method follows the equations:
\begin{align}
\alpha^\ast(t)  &= \alpha^\ast_0  + \mu_{\alpha^\ast}(t - t_0), \\
\delta(t)    &= \delta_0    + \mu_{\delta}(t - t_0),
\end{align}
where $(\alpha^\ast, \delta)$ is the binned position at time $t$, $(\alpha^\ast_0, \delta_0)$ are the fitted positions at the reference epoch $t_0$ (defined as the error-weighted mean time of the observations), and $(\mu_{\alpha^\ast}, \mu_{\delta})$ are the PMs. The linear fits were weighted by $1/\sigma^2_{\mathrm{bin}}$, and PM uncertainties were derived from the covariance matrix of the weighted least-squares solution. Note that $\alpha^\ast = \alpha\cos\overline{\delta}$, where $\overline{\delta}$ is the error-weighted mean declination of each individual star over its observations. This approach remains valid across the full DECam footprint, although projection effects should be considered in a full-field kinematic analysis.

After computing PMs for all stars, we cross-matched our catalog with the extinction-corrected photometric catalog from \citet{saha+19}. The final catalog contains $48\,239$~stars in the S10 detector field, which corresponds to the N10 detector field in the DECaPS-East observations. For this catalog, the median PM uncertainty for stars with $r_0<18$ is 0.65~mas~yr$^{-1}$. The full PM catalog, covering the complete DECam footprint, will be made publicly available with the forthcoming full-field analysis.

Finally, we transformed our PMs from equatorial ($\mu_{\alpha^\ast}$,$\mu_\delta$) into Galactic coordinates ($\mu^\ast_\ell,\mu_b$) following Eq.~6 of \citet{poleski13}. Since this transformation is a rotation by an angle that depends only on the position of each star, the uncertainties in Galactic coordinates were obtained by propagating the equatorial uncertainties through the same rotation.

\section{Comparison with \textit{Gaia}\,DR3 and VIRAC2} \label{sec:gaia}

\begin{figure*}
   \centering
   \includegraphics[width=0.7425\hsize]{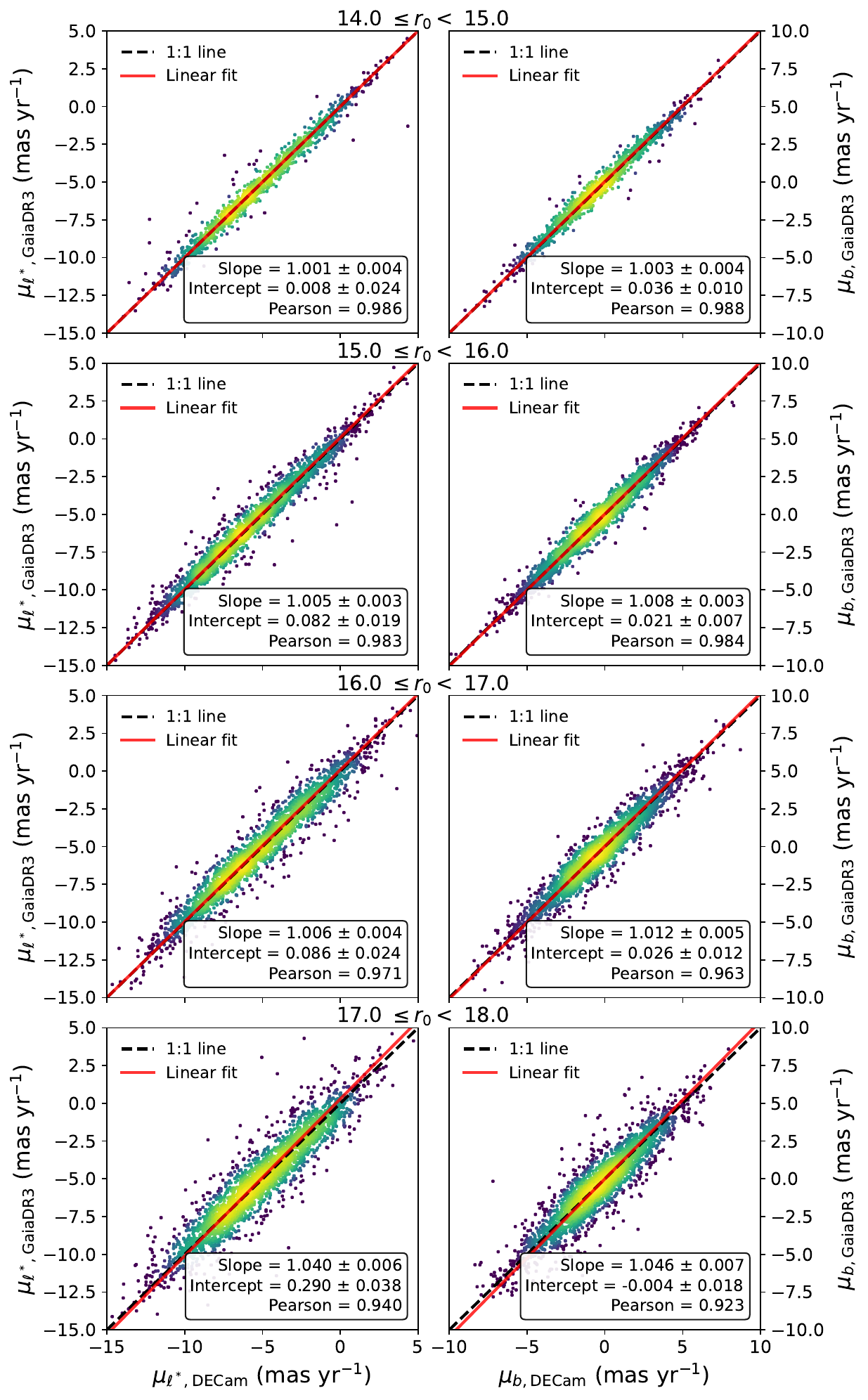}
   \caption{Direct comparison between \textit{Gaia}\,DR3 PMs ($\mu_{\ell^\ast,\mathrm{\textit{Gaia}\,DR3}}$, $\mu_{b,\mathrm{\textit{Gaia}\,DR3}}$) on the vertical axis and our DECam PMs ($\mu_{\ell^\ast,\mathrm{DECam}}$, $\mu_{b,\mathrm{DECam}}$) on the horizontal axis for different $r_0$ magnitude ranges. From top to bottom, the panels show magnitude bins: $14 \leq r_0 < 15$, $15 \leq r_0 < 16$, $16 \leq r_0 < 17$, and $17 \leq r_0 < 18$. The left column shows the $\mu_{\ell^\ast}$ comparison, while the right column shows the $\mu_b$ comparison. In each panel, the dashed black line represents the 1:1 relation, and the solid red line shows a linear fit. The slope and the intercept of each fit, in addition to Pearson's correlation coefficient, are indicated.}
   \label{Fig:gaia_comp}
\end{figure*}

\begin{figure}
   \centering
   \includegraphics[width=\hsize]{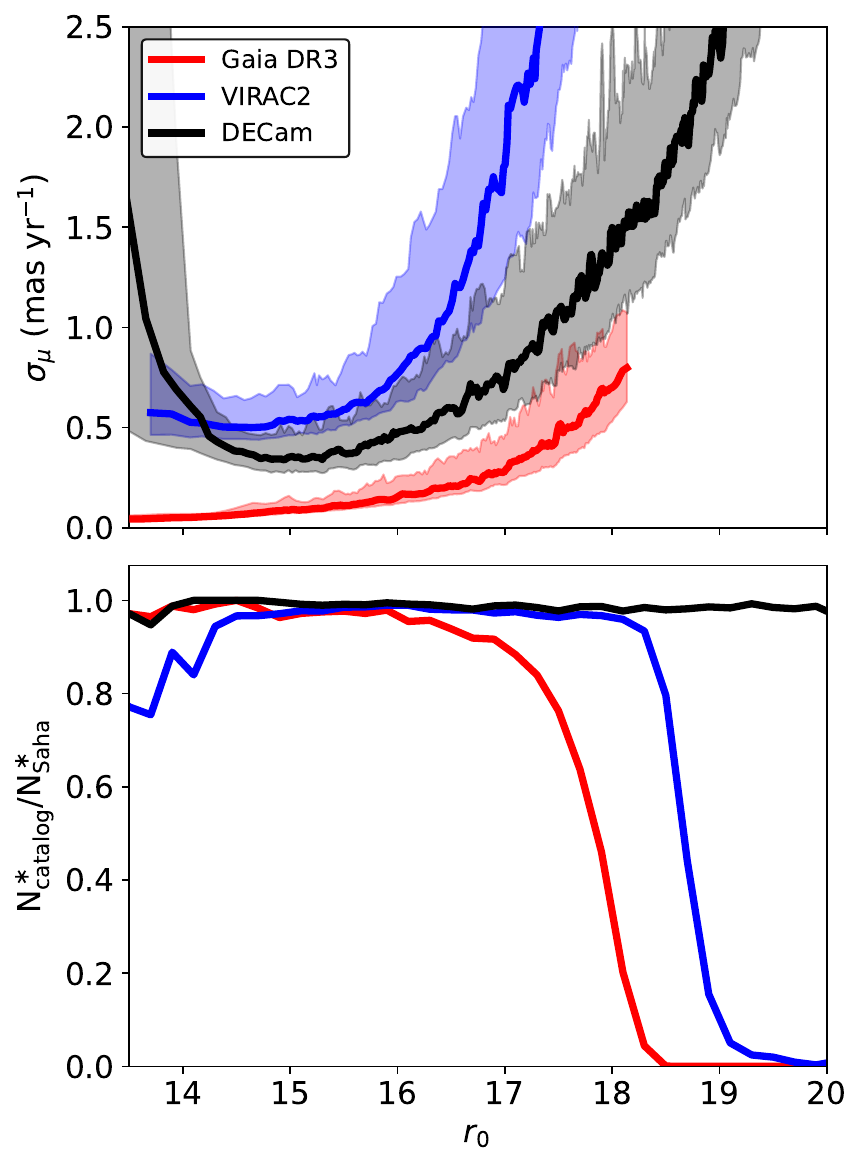}
   \caption{\textit{Top panel:} Total PM error, $\sigma_{\mu}$, as a function of $r_0$ magnitude for all stars in the region covered by the S10/N10 detectors, for our DECam catalog (black), \textit{Gaia}\,DR3 (red), and VIRAC2 (blue). The solid lines show the running median of $\sigma_{\mu}$, with the shaded regions indicating the 16th and 84th percentiles. \textit{Bottom panel:} Relative completeness of stars with available PM measurements as a function of $r_0$ (see text for definition).}
   \label{Fig:pms_error_comp}
\end{figure}

To test the precision of our results, we compared our DECam PM catalog with the \textit{Gaia}\,DR3 and VIRAC2 catalogs, the two most extensive PM catalogs currently available for the MW. \textit{Gaia}\,DR3 represents the state-of-the-art in astrometric precision, derived from space-based observations. On the other hand, VIRAC2 is the most extensive PM catalog of the Galactic bulge derived from ground-based observations, making it the most direct comparison for our DECam PMs. We performed the comparison within the same region covered by our pilot field.

\subsection{Proper motion measurements}\label{subsec:gaia_comp}

We compared our DECam PMs with those from \textit{Gaia}\,DR3 only for stars in common between our DECam catalog and \textit{Gaia}\,DR3, using the same cross-matching criteria described in Sect.~\ref{sec:astrometric_calibration}. Since we calibrated our astrometry using \textit{Gaia}\,DR3 as the absolute reference frame, our PMs are by construction in the same reference system as \textit{Gaia}\,DR3, facilitating a direct comparison between both datasets.

Fig.~\ref{Fig:gaia_comp} presents a direct one-to-one comparison between \textit{Gaia}\,DR3 PMs (vertical axis) and our DECam measurements (horizontal axis) across four magnitude bins from $14\leq r_0 < 15$ to $17\leq r_0 < 18$. For this comparison, we selected only stars with DECam PM uncertainties $\sigma_{\mu} < 1.0$~mas~yr$^{-1}$ to ensure high-quality measurements, and we omitted stars with $r_0 < 14$ and $r_0 > 18$ due to insufficient statistics, as they lie near the saturation limit of DECam and the magnitude limit of \textit{Gaia}\,DR3, respectively. 

The excellent agreement is evident from the distributions closely following the 1:1 relation (black dashed lines). For all magnitude bins, we performed a linear fit using an orthogonal distance regression to account for the uncertainties from both catalogs (red solid lines). The $\mu_{\ell^\ast}$ and $\mu_b$ comparisons both maintain strong agreement with \textit{Gaia}\,DR3, with no significant trend with magnitude, slopes close to unity and Pearson correlation coefficients above 0.9. The only exception is the faintest bin, where the slopes are $1.040\pm0.006$ and $1.046\pm0.007$ for $\mu_{\ell^\ast}$ and $\mu_b$, respectively. Although the deviation from unity is small, it amounts to ${\sim}6.5\sigma$ given the uncertainties of the fit, which could be attributed to the increasing astrometric uncertainties at fainter magnitudes in both catalogs. Furthermore, following the procedure of \citet{luna+23}, we analyzed the PM differences with respect to \textit{Gaia}\,DR3 normalized by the combined uncertainties, finding that both catalogs are consistent at the ${\sim}1.4\sigma$ level. We discuss this further in Appendix~\ref{app:inflation}.

The fitted intercepts are small in all cases, ranging from $0.008$ to $0.290$~mas~yr$^{-1}$ for $\mu_{\ell^\ast}$ and from $-0.004$ to $0.036$~mas~yr$^{-1}$ for $\mu_b$. The largest value corresponds to the faintest bin in $\mu_{\ell^\ast}$ and does not represent a genuine offset, since the fit pivots around the centroid of the distribution and the intercept compensates the slope slightly above unity discussed above. Evaluated at the median $\mu_{\ell^\ast}$ of the sample, the fit departs from the 1:1 relation by less than $0.1$~mas~yr$^{-1}$, negligible compared to the median PM uncertainty of 0.65~mas~yr$^{-1}$ up to $r_0=18$.

The near-zero intercepts and slopes close to unity confirm the success of our astrometric calibration procedure, indicating no significant systematic offset between the two catalogs. A comparison with VIRAC2 in the same format is presented in Appendix~\ref{app:virac2}.

\subsection{Proper motion uncertainties}

The top panel of Fig.~\ref{Fig:pms_error_comp} shows the total PM uncertainties ($\sigma_{\mu} = \sqrt{\sigma_{\mu_{\alpha^\ast}}^2 + \sigma_{\mu_{\delta}}^2}$) as a function of $r_0$ magnitude for all stars with available PM measurements in each catalog within the region covered by the S10/N10 detectors. The solid lines represent the running median uncertainties for our DECam catalog, \textit{Gaia}\,DR3, and VIRAC2, with shaded regions indicating the 16th and 84th percentiles.

As expected, \textit{Gaia}\,DR3 provides better PM precision across most of the magnitude range, with median uncertainties of 0.13~mas~yr$^{-1}$ for $r_0 < 17$ and 0.23~mas~yr$^{-1}$ for $r_0 < 18$. We do not expect to achieve better precision than \textit{Gaia}\,DR3, given that DECam is a ground-based facility and is therefore affected by atmospheric seeing, which broadens the PSF and degrades the centroiding precision. The median seeing of our images is ${\sim}1\arcsec$, well above the DECam pixel scale of $0.26\arcsec$~px$^{-1}$, so the centroiding precision is set by the seeing rather than by the sampling. In contrast, \textit{Gaia} observes from space, with an angular resolution of 0.059~\arcsec~px$^{-1}$ in the along-scan direction and 0.177~\arcsec~px$^{-1}$ in the across-scan direction. However, our DECam PMs show comparable precision to \textit{Gaia}\,DR3 at magnitudes approaching \textit{Gaia}'s magnitude limit ($r_0 \gtrsim 17$), demonstrating the complementary nature of the catalogs.

Compared to VIRAC2 \citep{smith2025}, the most extensive ground-based PM catalog covering the Galactic bulge, our DECam measurements show lower uncertainties across most of the magnitude range. The median uncertainties for our catalog are 0.44~mas~yr$^{-1}$ ($r_0 < 17$) and 0.65~mas~yr$^{-1}$ ($r_0 < 18$), compared to 0.67~mas~yr$^{-1}$ and 1.32~mas~yr$^{-1}$ for VIRAC2, respectively. Near our saturation limit ($r_0 < 14$), the DECam uncertainties increase sharply and VIRAC2 shows lower uncertainties, which might be explained by the fact that we are approaching the bright end where our astrometry degrades due to saturation. The median seeing of our DECam images is comparable to that of the VVV/VVVX observations (${\sim}1\arcsec$). Instead, the finer pixel scale of DECam compared to VISTA (0.34~\arcsec~px$^{-1}$) provides a better sampling of the PSF, which may contribute to the improved astrometric precision. However, since our astrometric uncertainties are approximated from the photometric errors rather than, for example, calibrated with artificial stars (see Sect.~\ref{subsec:uncertainties}), they could potentially be underestimated. We explore this in Appendix~\ref{app:inflation}.

\subsection{Relative completeness}

A proper characterization of the completeness of our catalog would require artificial star tests, in which synthetic sources are injected into the images and recovered through the same detection and PM measurement pipeline. However, given the large volume of data in our catalog, such tests are computationally expensive and inefficient. We therefore opted to estimate a relative completeness with respect to the photometric catalog of \citet{saha+19}, which is one of the deepest catalogs of the region obtained with wide-field instruments.

The bottom panel of Fig.~\ref{Fig:pms_error_comp} shows the relative completeness of stars with available PM measurements as a function of $r_0$ magnitude, defined as $N^\ast_{\mathrm{catalog}}/N^\ast_{\mathrm{Saha}}$, where $N^\ast_{\mathrm{catalog}}$ is the number of sources with a measured PM in each catalog and $N^\ast_{\mathrm{Saha}}$ is the total number of sources in the complete photometric catalog of \citet{saha+19}, both counted per $r_0$ magnitude bin.

Our PM catalog recovers nearly all the sources detected by \citet{saha+19}, maintaining a relative completeness close to 100\% across the full magnitude range. In contrast, \textit{Gaia}\,DR3 reaches comparable completeness for bright stars but drops sharply around $r_0\approx18$, near its magnitude limit. VIRAC2 reaches nearly 100\% relative completeness over $14.5 \lesssim r_0 \lesssim 18.5$, decreasing at both the bright and faint ends, where our catalog extends ${\sim}1$~magnitude deeper. However, this comparison should be interpreted with caution, since our catalog includes the original observations of \citet{saha+19} and is therefore expected to be nearly complete by construction. In addition, we verified that most of the missing bright sources and many of the missing faint sources are present in the VIRAC2 rejected-source catalog, indicating that they were detected but did not satisfy the quality criteria described by \citet{smith2025}. We therefore consider the completeness of both catalogs to be comparable, with the two surveys being complementary.

\section{Galactic bulge and foreground disk stellar kinematics} \label{sec:kinematics}

\begin{figure*}
    \centering
    \includegraphics[width=\hsize]{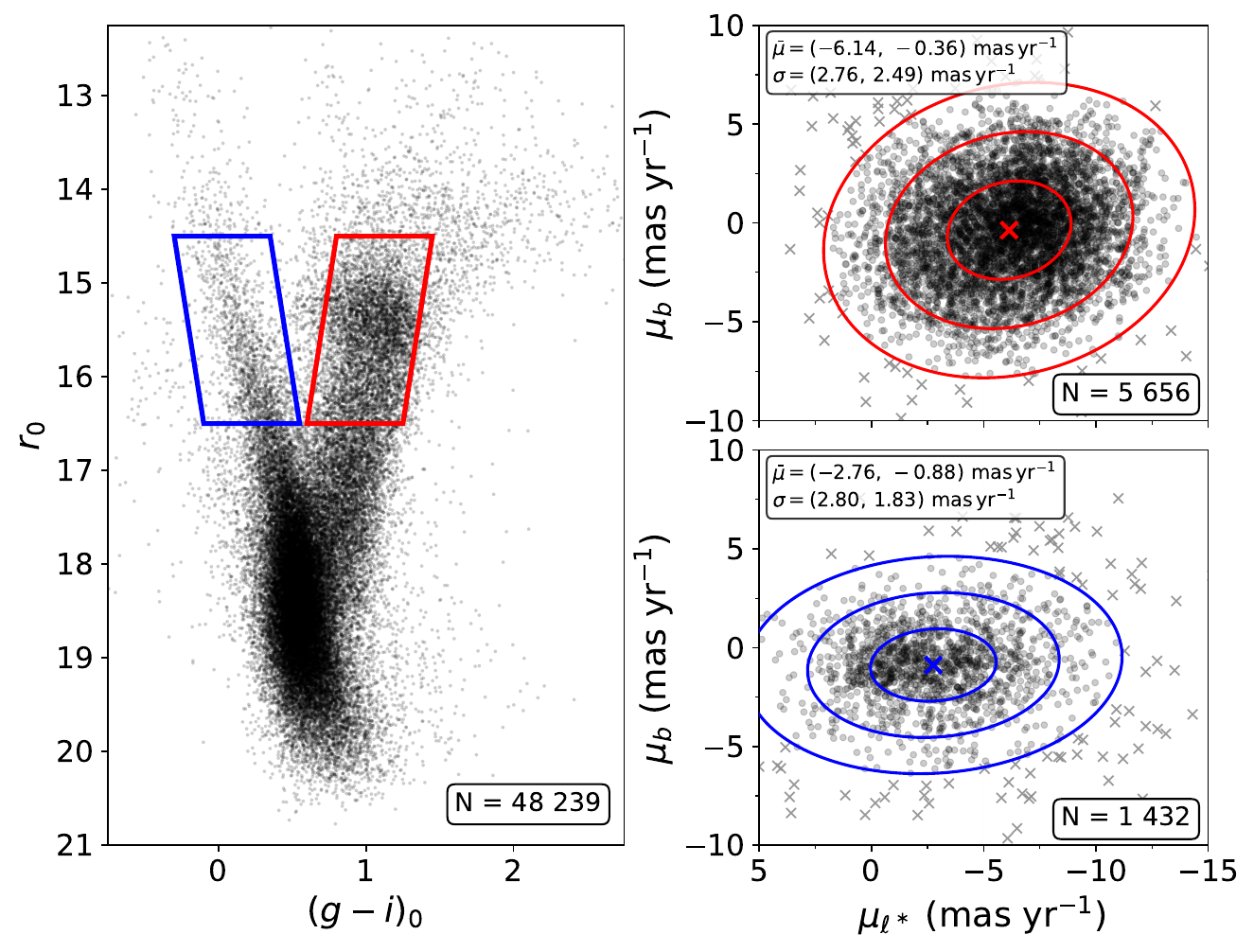}
    \caption{\textit{Left:} $r_0$, $(g-i)_0$ CMD for all $N = 48\,239$ stars in the S10/N10 field, based on the extinction-corrected photometric catalog of \citet{saha+19}. The blue and red boxes show the regions used to select the foreground disk and Galactic bulge RC and RGB stars, respectively. \textit{Right:} VPDs of the disk (\textit{bottom panel}) and bulge (\textit{top panel}) samples. The mean PM $\bar{\mu}$ and dispersions $\sigma$ of each Gaussian fit are indicated in each panel. Grey crosses indicate stars rejected by the $3\sigma$ clipping and excluded from the fit. Solid ellipses show the $1\sigma$, $2\sigma$, and $3\sigma$ contours of each fit, with the cross marking their centroids.}
    \label{Fig:cmd_samples_vpd}
\end{figure*}

The bulge and disk sequences can be well separated at magnitudes brighter than the bulge MSTO, as shown by \citet{clarkson+08}; therefore, we can estimate the kinematics of both components by isolating photometrically selected samples. To this end, we consider two boxes of the same size in the CMD (see Fig.~\ref{Fig:cmd_samples_vpd}), where the red one contains mostly red clump (RC) and red giant branch (RGB) stars of the Galactic bulge, and the blue one mostly contains foreground disk stars. Both boxes are restricted to magnitudes brighter than $r_0 \sim 16.5$, where the two sequences are well separated in color, and thus the contamination is lower. This results in a bulge sample of $5\,656$ stars and a disk one of $1\,432$.

For each sample, we derive the kinematics by fitting a single bivariate Gaussian to the PM distribution (Fig.~\ref{Fig:cmd_samples_vpd}), accounting for individual PM uncertainties through extreme deconvolution to recover the intrinsic velocity dispersions. By looking at the vector point diagram (VPD), it is clear that the two components display distinct kinematics, with the bulge sample shifted towards more negative $\mu_{\ell^\ast}$ and showing a larger spread, particularly in $\mu_b$. Since both samples, despite the photometric selection, retain some contamination between components, we first apply an iterative $3\sigma$ clipping that rejects the outliers of each distribution (grey crosses in Fig.~\ref{Fig:cmd_samples_vpd}). The resulting kinematic parameters for each component are summarized in Table~\ref{Tab:gmm} and discussed below. Uncertainties on all parameters were estimated by bootstrap.

For the bulge sample, we obtain mean PMs of $\bar{\mu}_{\ell^\ast} = -6.14 \pm 0.04$~mas~yr$^{-1}$ and $\bar{\mu}_{b} = -0.36 \pm 0.04$~mas~yr$^{-1}$, with velocity dispersions of $\sigma_{\mu_{\ell^\ast}} = 2.76 \pm 0.03$~mas~yr$^{-1}$ and $\sigma_{\mu_{b}} = 2.49 \pm 0.02$~mas~yr$^{-1}$. These dispersions fall well within the ranges found in previous studies for the Baade's window and nearby fields, where $\sigma_{\mu_{\ell^\ast}}$ ranges from 2.5~mas~yr$^{-1}$ to 3.1~mas~yr$^{-1}$ and $\sigma_{\mu_{b}}$ ranges from 2.4~mas~yr$^{-1}$ to 3.0~mas~yr$^{-1}$ depending on the study \citep[e.g.,][]{kuijken+02,kozlowski+06, clarkson+08, clarke+19, terry+20}. Moreover, we found a velocity anisotropy of $\sigma_{\mu_{\ell^\ast}}/\sigma_{\mu_b} = 1.11 \pm 0.02$, which also agrees with the values reported in the studies cited above.

For the foreground disk component, we obtain mean PMs of $\bar{\mu}_{\ell^\ast} = -2.76 \pm 0.08$~mas~yr$^{-1}$ and $\bar{\mu}_{b} = -0.88 \pm 0.06$~mas~yr$^{-1}$, with velocity dispersions of $\sigma_{\mu_{\ell^\ast}} = 2.80 \pm 0.06$~mas~yr$^{-1}$ and $\sigma_{\mu_{b}} = 1.83 \pm 0.05$~mas~yr$^{-1}$, resulting in an anisotropy of $\sigma_{\mu_{\ell^\ast}}/\sigma_{\mu_b} = 1.53 \pm 0.05$, in good agreement with the values reported for the foreground disk in previous studies \citep[e.g.,][]{calamida+2014, terry+20}.

We verified that for the bulge sample, the mean PMs and the anisotropy remain almost unchanged when we omit the 3$\sigma$ clipping, with the velocity dispersions increasing only marginally ($\lesssim$~3\%). However, for the foreground disk sample, the effect is larger. Without clipping, the dispersions increase by ${\sim}15\%$ in $\sigma_{\mu_{\ell^\ast}}$ and ${\sim}30\%$ in $\sigma_{\mu_b}$. This could be explained by the fact that the disk PM distribution is more strongly affected by nearby foreground stars, which have intrinsically larger PMs, increasing the dispersions. In any case, both sets of values remain consistent with the literature, but the $3\sigma$-clipped values agree better with previous studies, so we adopt them throughout.

\section{Cleaning the Galactic bulge stellar population} \label{sec:bulge_decontamination}

    \begin{figure*}
        \centering
        \includegraphics[width=\hsize]{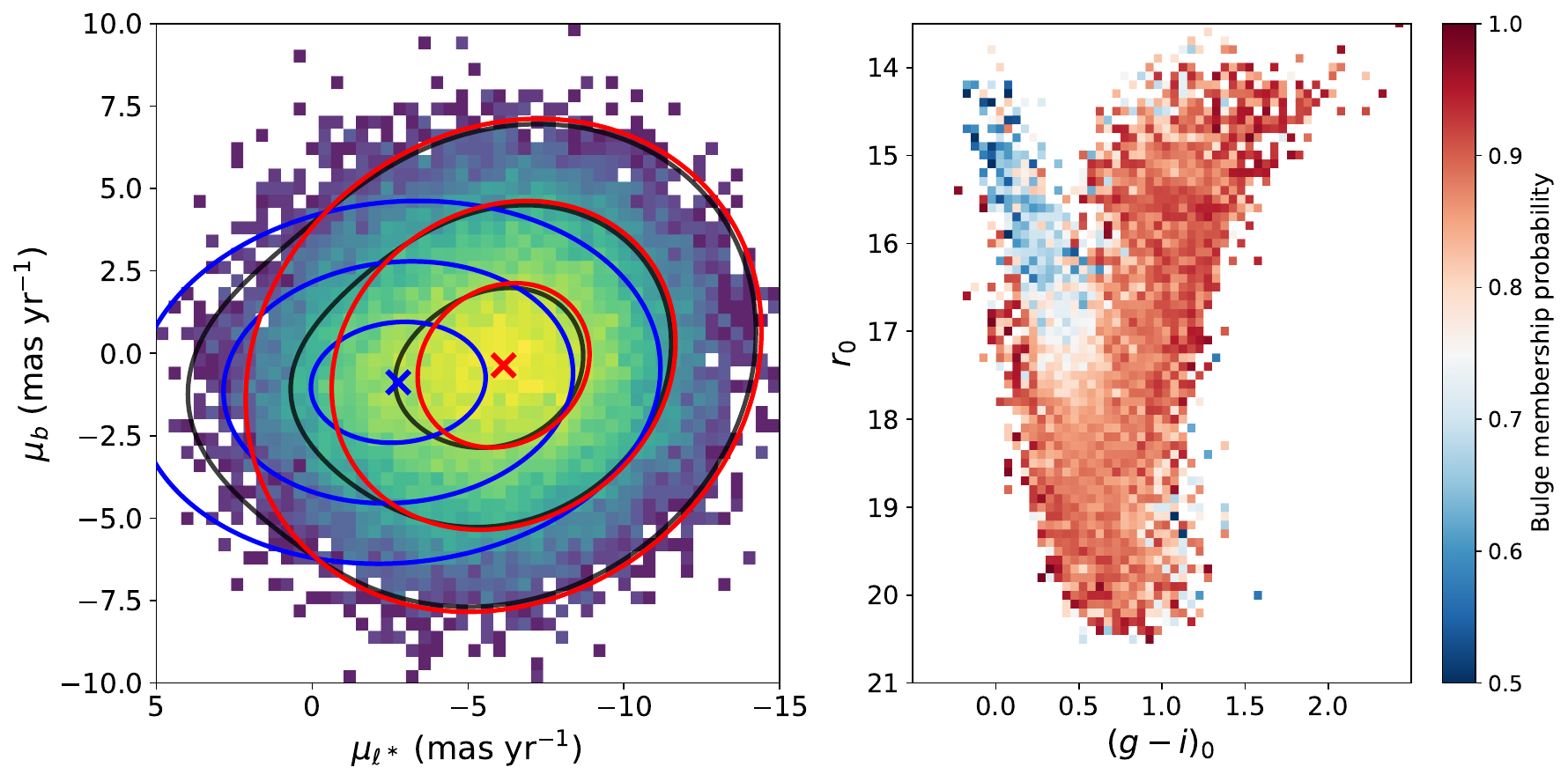}
        \caption{\textit{Left:} PM distribution in the S10/N10 field, colored by stellar density. The red and blue ellipses show the 1.0, 2.0, and 3.0$\sigma$ confidence contours of the Galactic bulge and foreground disk components of the GMM, respectively, with crosses marking their centroids. Black contours show the combined GMM at the same confidence levels. The GMM parameters are listed in Table~\ref{Tab:gmm}. \textit{Right:} $r_0$, $(g-i)_0$ CMD, with each bin colored by the mean bulge membership probability, where redder bins indicate higher values and bluer bins lower values.}
        \label{Fig:gmm_vpd}
    \end{figure*}

\begin{table}[ht]
    \caption{Parameters of the best-fit GMM for the PM distribution.}
    \label{Tab:gmm}
    \centering
    {
    \begin{tabular}{lcc}
        \hline\hline \noalign{\smallskip}
        Parameter & Galactic bulge & Foreground disk \\
        \noalign{\smallskip}\hline\noalign{\smallskip}
        Weight & $0.856 \pm 0.004$ & $0.144 \pm 0.004$ \\
        $\bar{\mu}_{\ell^\ast}$ (mas~yr$^{-1}$) & $-6.14 \pm 0.04$ & $-2.76 \pm 0.08$ \\
        $\bar{\mu}_{b}$ (mas~yr$^{-1}$) & $-0.36 \pm 0.04$ & $-0.88 \pm 0.06$ \\
        $\sigma_{\mu_{\ell^\ast}}$ (mas~yr$^{-1}$) & $2.76 \pm 0.03$ & $2.80 \pm 0.06$ \\
        $\sigma_{\mu_{b}}$ (mas~yr$^{-1}$) & $2.49 \pm 0.02$ & $1.83 \pm 0.05$ \\
        $\sigma_{\mu_{\ell^\ast}}/\sigma_{\mu_{b}}$ & $1.11 \pm 0.02$ & $1.53 \pm 0.05$ \\
        \noalign{\smallskip}\hline
    \end{tabular}
    }
    \tablefoot{The table shows the weight, means ($\bar{\mu}_{\ell^\ast}$, $\bar{\mu}_{b}$), velocity dispersions ($\sigma_{\mu_{\ell^\ast}}$, $\sigma_{\mu_{b}}$), and anisotropy ratios ($\sigma_{\mu_{\ell^\ast}}/\sigma_{\mu_{b}}$) of each component, with their $1\sigma$ uncertainties estimated by bootstrap.}
\end{table}

One of the most powerful applications of PMs is their ability to distinguish different kinematic components of our Galaxy. In particular, PMs enable the separation of Galactic bulge stars from foreground contaminants, which is crucial for the detailed analysis of the properties of the different stellar components in highly crowded fields. In this section, we isolate the Galactic bulge stellar population by applying a strict cut in $\mu_{\ell^\ast}$, the coordinate along which the bulge and the foreground disk are best separated (see Sect.~\ref{sec:kinematics}). We use a two-component GMM to estimate the global contamination of the field, and we further refine the selection using \textit{Gaia}\,DR3 parallaxes, when available. This procedure aims to produce a clean CMD of Galactic bulge stars reaching down to magnitudes fainter than the MSTO.

\subsection{Gaussian mixture model from DECam PMs}\label{sec:decam_clean}

To estimate the global contamination of the field, we modeled the PM distribution in the VPD with a Gaussian Mixture Model (GMM), using the \textsc{GaussianMixture} routines from the \textsc{Python} package \textsc{scikit-learn} \citep{pedregosa2011}. We use only two Gaussian components, one for the Galactic bulge and one for the foreground disk, since the contribution of other populations along this line of sight is negligible (e.g., the halo contribution in the Baade's window is estimated to be around $0.1\%$ by \citealt{hill+2011}). Rather than fitting all the parameters, we fix the means and dispersions of each component to the values derived from the photometrically selected samples in Sect.~\ref{sec:kinematics}, and fit only the relative weights of the two components. The best-fit model, shown in the left panel of Fig.~\ref{Fig:gmm_vpd}, is dominated by the bulge component, which accounts for a fraction $w_{\rm bulge} = 0.856\pm0.004$ of the field, while the foreground disk component accounts for the remaining $w_\mathrm{disk} = 0.144\pm0.004$. The parameters for each Gaussian component are listed in Table~\ref{Tab:gmm}.

The fitted weight of the foreground component provides a direct estimate of the overall foreground contamination in the field. Prior to any kinematic selection, about $14.4\%$ of the stars might belong to the foreground disk population. This is broadly consistent with the ${\sim}14\%$ contamination found by \citet{clarkson+08} in a smaller, deeper HST field in SWEEPS, and with the ${\sim}10\%$ estimated by \citet{hill+2011} from a Besançon model \citep{robin+2003,picaud+2004} in the Baade's window.

From the best GMM, the bulge membership probability ($P_\mathrm{bulge}$) for each star was computed using Bayes' theorem. For a star with ($\mu_{\ell^\ast},\mu_b$), $P_\mathrm{bulge}$ is given by:

\begin{equation}
    P_\mathrm{bulge}(\mu_{\ell^\ast},\mu_b)=\frac{
        w_\mathrm{bulge} \,\mathcal{N}_\mathrm{bulge}(\mu_{\ell^\ast}, \mu_b)
    }{
        w_\mathrm{bulge} \,\mathcal{N}_\mathrm{bulge}(\mu_{\ell^\ast}, \mu_b)+w_\mathrm{disk} \,\mathcal{N}_\mathrm{disk}(\mu_{\ell^\ast}, \mu_b)
    },
    \label{eq:membership_probability}
\end{equation}

\noindent where $\mathcal{N}_\mathrm{component}(\mu_{\ell^\ast}, \mu_b)$ denotes the bivariate Gaussian distribution for the respective component, evaluated at the observed PM.

\begin{figure*}
    \centering
    \includegraphics[width=\hsize]{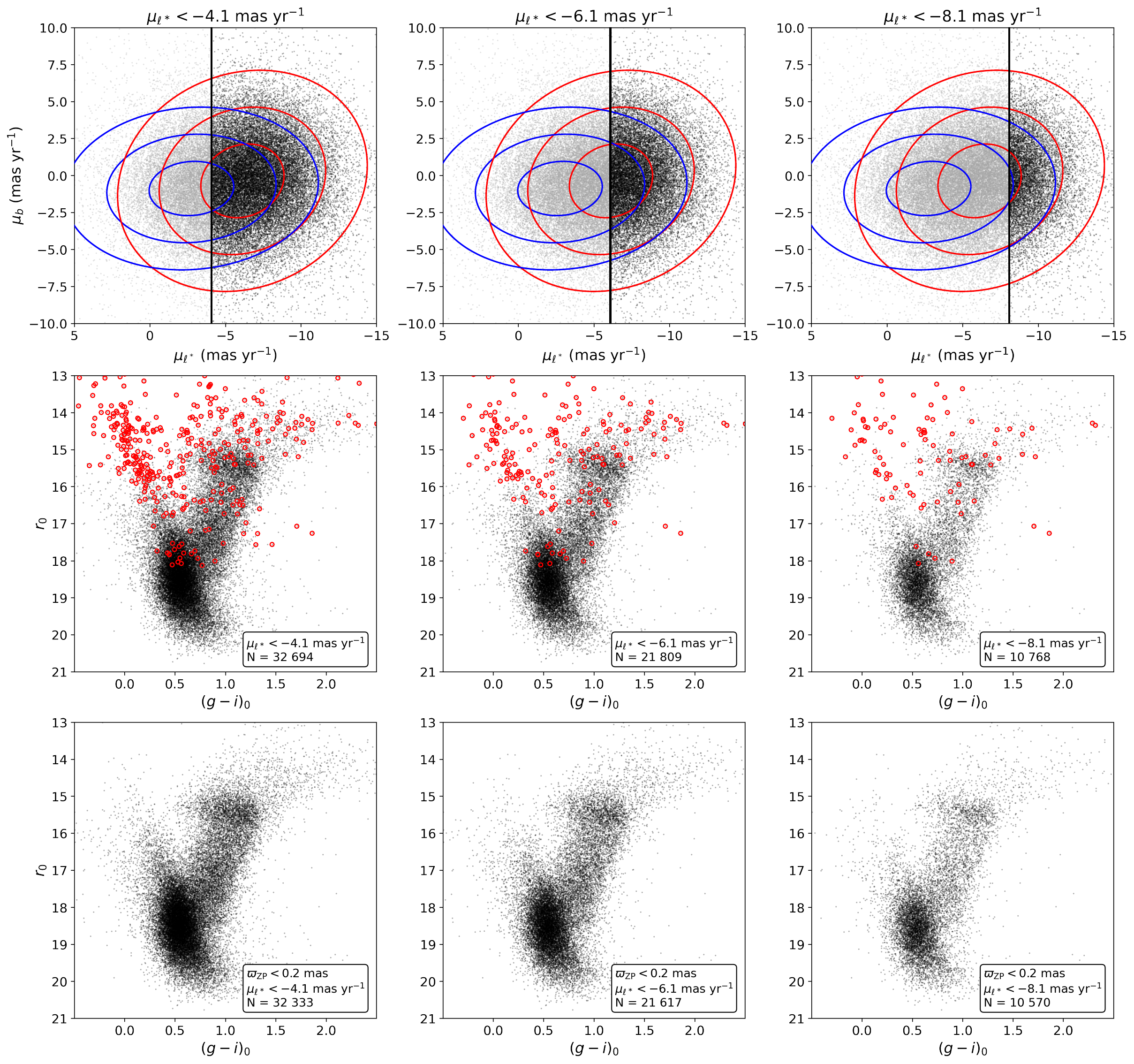}
    \caption{Kinematic decontamination of the Galactic bulge population for three different cuts in $\mu_{\ell^\ast}$ (one per column). \textit{Top rows:} VPDs of the field with stars selected by each cut (red vertical line) colored by density, and rejected stars shown in grey. The red and blue ellipses show the same confidence regions of the left panel of Fig.\ref{Fig:gmm_vpd}. \textit{Middle rows:} CMD of the stars satisfying each PM cut. Red open circles show the stars subsequently removed by the \textit{Gaia} parallax cut. \textit{Bottom rows:} CMD of the final sample, after applying both the PM and the parallax cut. The number of selected stars are indicated in each panel.}
    \label{Fig:cmd_mulcuts}
\end{figure*}

The right panel of Fig.~\ref{Fig:gmm_vpd} shows the extinction-corrected CMD colored by the mean $P_{\rm bulge}$ per bin. Since the bulge component largely dominates the field ($w_{\rm bulge} > 0.85$) and both components strongly overlap in the VPD, the GMM is biased toward the bulge, so that even genuine foreground disk stars are often assigned $P_{\rm bulge} > 0.5$. For this reason, for a better visualization, we truncated the color scale at $P_\mathrm{bulge}=0.5$. Despite this, as expected, the bulge RC, RGB, and the region below the MSTO display the highest probabilities, whereas the foreground disk exhibits a well-defined lower-probability sequence. Our probabilities are also consistent with the values obtained by \citet{marchetti+22} from \textit{Gaia}\,DR3 PMs and BDBS photometry in the Baade's window (see their Fig.~8).

\subsection{Kinematic cleaning of the bulge population}

\begin{figure}
    \centering
    \includegraphics[width=\hsize]{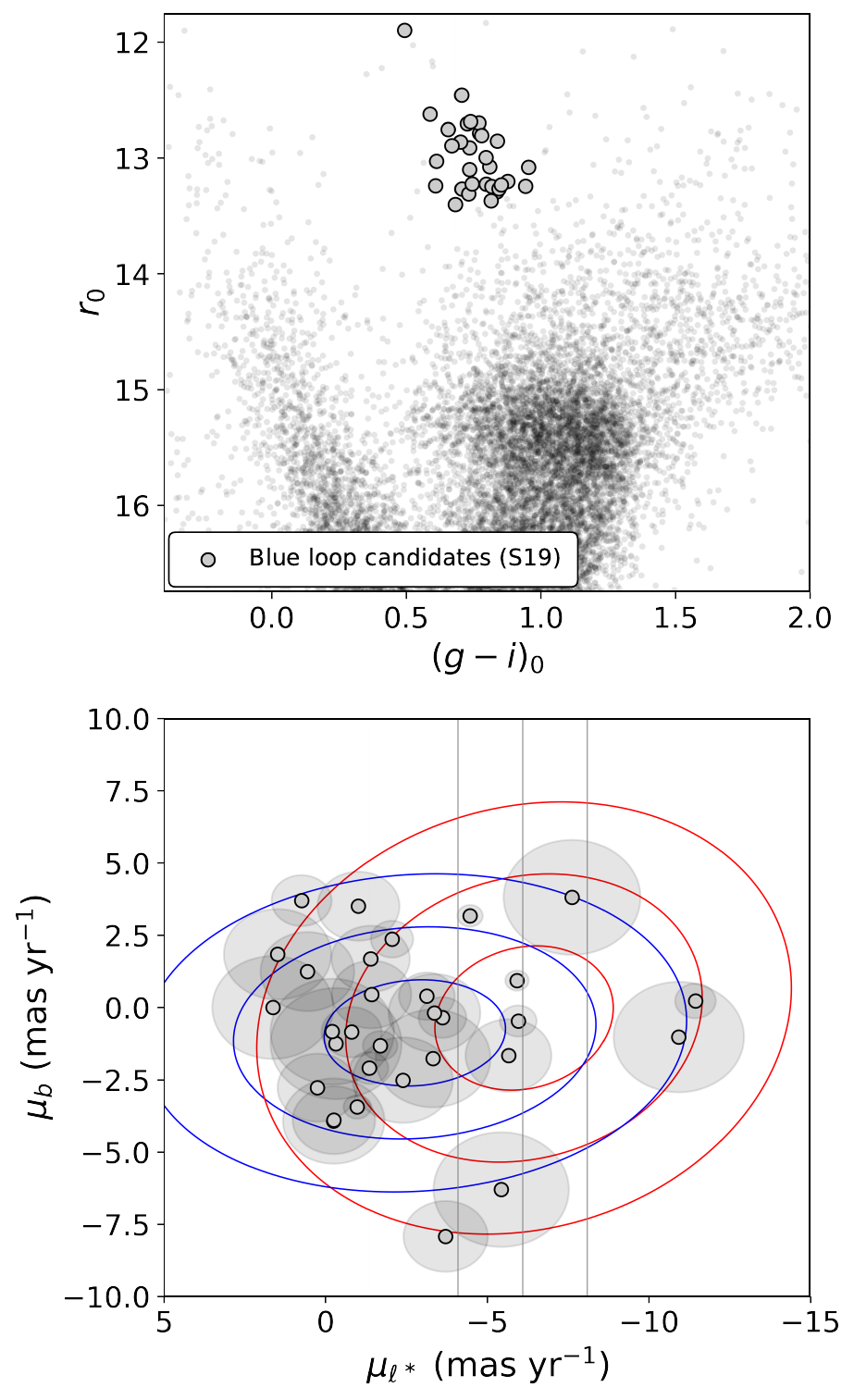}
    \caption{Candidate blue loop stars identified by \citet{saha+19} in the pilot field. \textit{Top panel:} $r_0$, $(g-i)_0$ CMD of the candidates with a total PM uncertainty below 3~mas~yr$^{-1}$ (grey circles), over the CMD of the full field (black points). \textit{Bottom panel:} PMs of the same stars, with the individual $1\sigma$ error ellipses shown in grey. The red and blue ellipses show the $1\sigma$, $2\sigma$, and $3\sigma$ contours of the Galactic bulge and foreground disk components, and the grey vertical lines mark the three $\mu_{\ell^\ast}$ cuts adopted, $\mu_{\ell^\ast}^\mathrm{cut}=-4.1,-6.1,$ and $-8.1$~mas~yr$^{-1}$.}
    \label{Fig:blue_loops}
\end{figure}

The ability of PMs to separate the bulge from the foreground disk along $\mu_{\ell^*}$ is well known \citep[e.g.,][]{kuijken+02,clarkson+08, calamida+2014, terry+20}. As shown in the previous sections, our DECam PMs successfully reproduce this separation, which allows us to isolate the bulge population through a strict cut in $\mu_{\ell^*}$, rather than relying on the GMM membership probabilities $P_{\rm bulge}$. We adopted three different cuts, $\mu_{\ell^*}^\mathrm{cut}=-4.1, -6.1,$ and $-8.1$~mas~yr$^{-1}$, by requiring $\mu_{\ell^*}<\mu_{\ell^*}^\mathrm{cut}$. In particular, the central cut, $\mu_{\ell^*}^\mathrm{cut}=-6.1$~mas~yr$^{-1}$, coincides with the mean PM of the bulge component, while the strictest cut, $\mu_{\ell^*}^\mathrm{cut}=-8.1$~mas~yr$^{-1}$, is consistent with the $\mu_{\ell^\ast}<-2.0$~mas~yr$^{-1}$ cut adopted by \citet{clarkson+08} and most of the literature, but tied to the \textit{Gaia}\,DR3 astrometric system.

The selections according to the three adopted cuts are illustrated in Fig.~\ref{Fig:cmd_mulcuts}. The top row shows the selected region in the VPD, while the middle row shows the corresponding CMD of the retained stars. It is evident that as the cut becomes more stringent, the foreground blue main sequence is progressively suppressed. However, given the significant overlap between the bulge and disk PM distributions, a non-negligible number of stars remains in the blue sequence even for the strictest cut (see middle right panel of Fig.~\ref{Fig:cmd_mulcuts}).

To further refine the bulge samples and remove the clear foreground disk contaminants that survived the kinematic selection, we applied an additional criterion based on \textit{Gaia}\,DR3 parallaxes, when available. First, we generated zero-point offset corrected parallaxes \citep{lindegren+2021}, hereafter $\varpi_\mathrm{ZP}$, and retained only measurements with RUWE $< 1.4$ and $\varpi_{\mathrm{ZP}}/\sigma_\varpi > 5$ to ensure reliability. Only 1\,056 stars (2.2\% of the total) have parallaxes satisfying these quality criteria. Then, following \citet{marchetti+22}, we considered as foreground contaminants all stars with $\varpi_{\mathrm{ZP}} > 0.2$ mas, corresponding to heliocentric distances closer than 5 kpc. This removes 961 stars (2\% of the total) from the full catalog.

In the middle row of Fig.~\ref{Fig:cmd_mulcuts}, the stars that survive each $\mu_{\ell^\ast}$ cut but are subsequently removed by the parallax criterion are marked with red open circles. For the three cuts, at least half of the parallax-removed stars fall in the blue sequence, around $(g-i)_0<0.6$ and $r_0<17$, confirming that the parallax cut preferentially removes nearby foreground disk stars missed by the PM selection. The bottom row shows the final samples, after applying both the $\mu_{\ell^\ast}$ and the $\varpi_\mathrm{ZP}$ cuts. We verified that the kinematic selection already removes the bulk of the foreground stars flagged by the parallax criterion, about 62\%, 80\%, and 90\% for $\mu_{\ell^\ast}^\mathrm{cut}=-4.1,-6.1,$ and $-8.1$~mas~yr$^{-1}$, respectively. Although the parallax cut removes only a small number of stars, these are concentrated in the disk sequence discussed above.

Some blue sequence stars remain in our final cleaned samples, even for the strictest cut. These likely represent residual inner disk contaminants with distances beyond our 5 kpc threshold, BSSs belonging to the bulge \citep[e.g.,][]{clarkson2011}, and/or blue horizontal branch stars in the bulge population \citep[e.g.,][]{peterson2001}. In any case, stars with $(g-i)_0 < 0.6$ and $r_0 < 17$ constitute only ${\sim}3.5\%$ of our final sample for the least restrictive cut, decreasing to ${\sim}2.5\%$ for the strictest one, indicating that the majority of the foreground contamination has been successfully removed.

Finally, we also examine in more detail the stars proposed by \citet{saha+19} as blue loop candidates, populating the region $0.5\lesssim(g-i)_0\lesssim1.0$ above the RC (see their Fig.~15). \citet{saha+19} noticed that the colors of these stars become tighter after de-reddening, unlike those of the foreground main sequence, and that a subset of them shows spectroscopic signatures consistent with the bulge population (see their Sects.~6.4 and 6.5), suggesting that these stars belong to the bulge. Therefore, we cross-matched the candidate list of \citet[][Saha, private communication]{saha+19} with our PM catalog, finding 39 of them in the pilot field. Since our PM uncertainties increase sharply at the brightest magnitudes (see Fig.~\ref{Fig:pms_error_comp}), we adopted a permissive cut and discarded only those stars with $\sigma_\mu > 3$~mas~yr$^{-1}$, leaving 31 stars with a mean uncertainty of 1.7~mas~yr$^{-1}$.

As shown in the top panel of Fig.~\ref{Fig:blue_loops}, these stars populate the region above the bulge RC in our CMD, forming the roughly vertical sequence identified by \citet{saha+19}. However, their kinematics are more consistent with the foreground disk component than with the Galactic bulge, with median values of $\mu_{\ell^\ast} = -2.71$~mas~yr$^{-1}$ and $\mu_b = -0.62$~mas~yr$^{-1}$ (bottom panel). Nevertheless, the GMM assigns these stars mean membership probabilities of 68\% for the bulge and 32\% for the foreground disk, reflecting the substantial overlap between the two kinematic components. We caution that these probabilities were derived from a magnitude-independent GMM, and a magnitude-dependent analysis could yield different estimates, since the relative fractions of foreground disk and bulge stars may vary with magnitude. However, the limited number of bright stars in our pilot field prevents a robust assessment of this effect, which we plan to investigate in our forthcoming full-field analysis. Despite this uncertainty in the membership probabilities, a similar trend was found by \citet{marchetti+22} for stars in this region of the CMD, using \textit{Gaia}\,EDR3 PMs (see their Fig.~9). \citet{rich+2020} have argued that these stars are foreground red giants: the PM result is consistent with that. Further, the \textit{Gaia}\,DR3 parallax results place our sample of 31 stars at a mean distance of 2.96~kpc from us; as a result, all of them were removed by the parallax cut. However the statistical errors (which correlate with the errors in PMs), as well as the Lutz-Kelker bias \citep{lutzkelker1973} are at play here, and the results are less than definitive. Also, of note, is that \citet{matsunaga+2011} have found Cepheids within 10 pc of the central black hole at the Galactic center, showing the presence of very young stars (albeit rare) towards the bulge region, presumably with kinematic properties that differ from the dominant stellar populations of the bulge. The controversy over the origin and nature of these objects, therefore, remains unresolved. We will re-address it when we present the forthcoming results from the full-field analysis.

The individual astrometric properties and GMM membership probabilities of the blue-loop candidates analyzed here are presented in Table~\ref{Tab:blue_loops}.

\subsection{Purity and completeness of the clean bulge sample}

\begin{figure}
    \centering
    \includegraphics[width=\hsize]{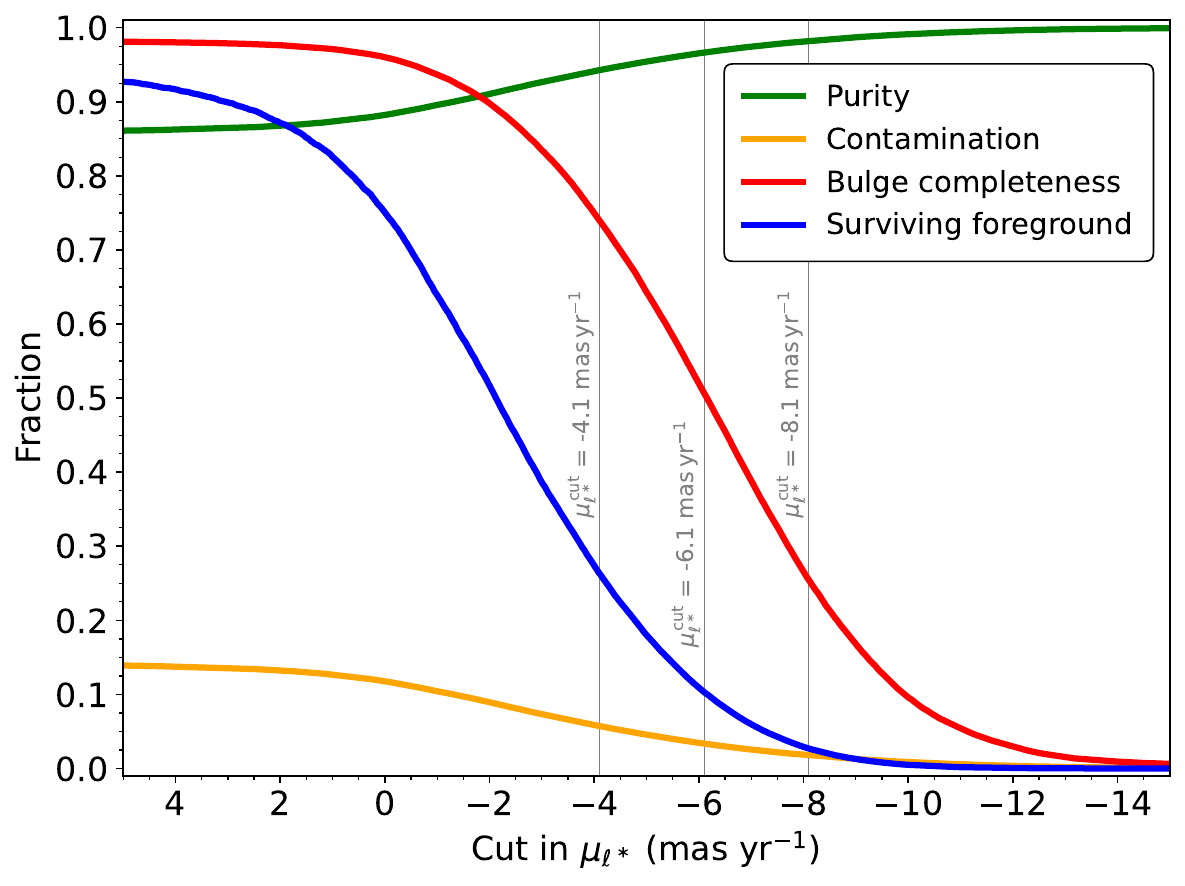}
    \caption{Purity (green), contamination (orange), bulge completeness (red), and surviving foreground (blue) of the selected sample as a function of the cut in $\mu_{\ell^\ast}$, estimated from the GMM membership probabilities. The vertical grey lines show the three adopted cuts, $\mu_{\ell^\ast}^\mathrm{cut}=-4.1,-6.1,$ and $-8.1$~mas~yr$^{-1}$.}
    \label{Fig:purity_completeness}
\end{figure}

\begin{figure}
    \centering
    \includegraphics[width=\hsize]{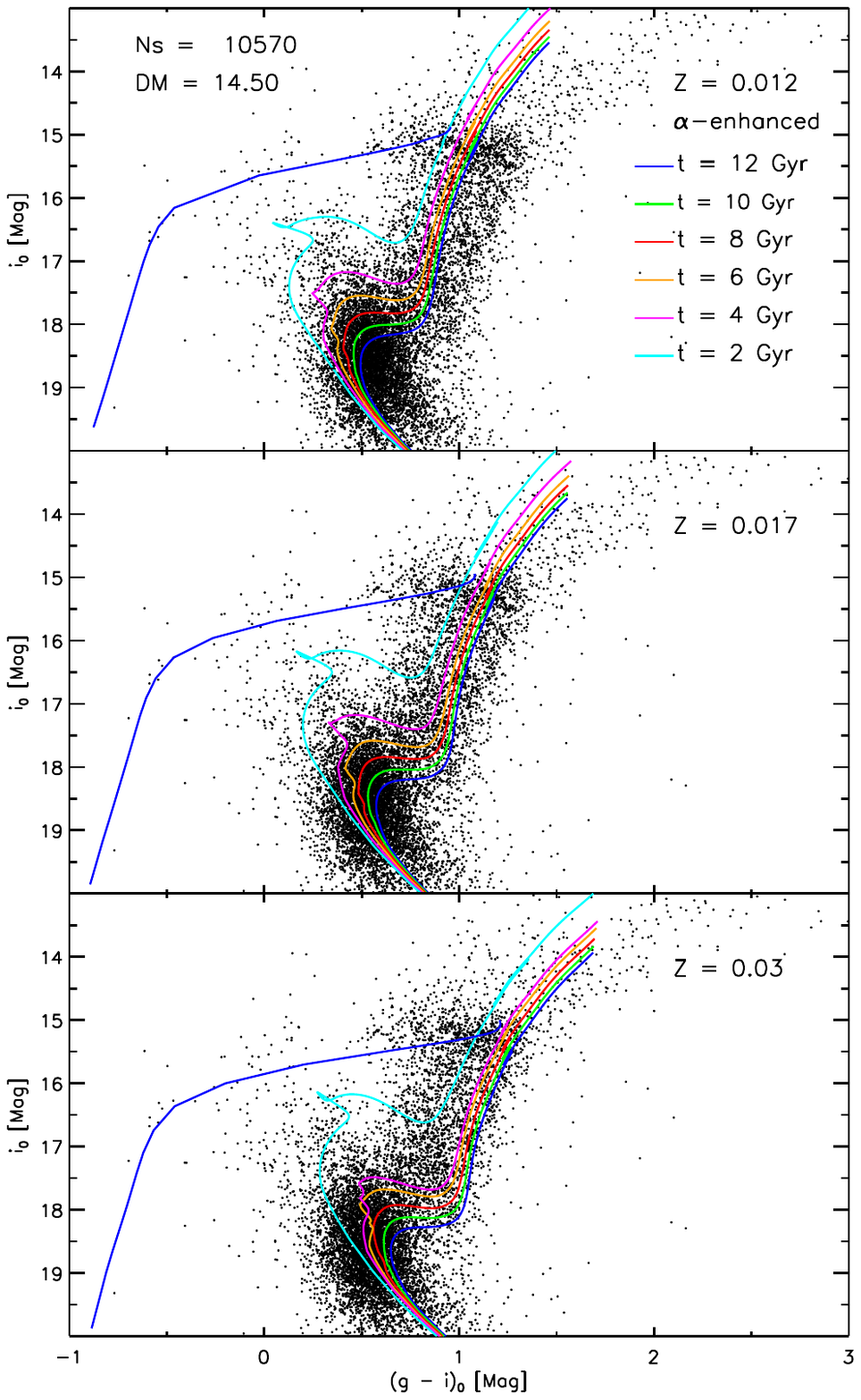}
    \caption{$i_0,\ (g-i)_0$ CMD of the bulge member stars selected in PM and parallax. Isochrones for different metallicities (sub-solar, top panel), solar (middle) and super-solar (bottom) and different ages, are over-plotted. The number of total stars (Ns) and the assumed distance modulus (DM) are also labeled.}
    \label{Fig:cmd_iso}
\end{figure}

We used the GMM membership probabilities to quantify the quality of our final bulge samples, after applying both the $\mu_{\ell^\ast}$ and the parallax cuts, as a function of the adopted kinematic cut. For a given selection, we define the purity as the expected fraction of bulge stars among the selected sources, calculated as their mean bulge membership probability, and the contamination as its complement. Similarly, we define the bulge completeness as the fraction of the total expected bulge population retained by the selection, estimated from the sum of the membership probabilities. The surviving foreground is defined analogously, using the disk membership probabilities. Fig.~\ref{Fig:purity_completeness} shows these quantities as a function of the cut in $\mu_{\ell^\ast}$.

There is an evident trade-off between purity and bulge completeness. As the cut becomes stricter, the purity increases and the surviving foreground drops sharply, but at the expense of a lower bulge completeness. For our three adopted cuts, the purity increases from ${\sim}94\%$ at $\mu_{\ell^\ast}^\mathrm{cut}=-4.1$~mas~yr$^{-1}$ to ${\sim}98\%$ at $\mu_{\ell^\ast}^\mathrm{cut}=-8.1$~mas~yr$^{-1}$, while the bulge completeness decreases from ${\sim}74\%$ to ${\sim}28\%$, respectively. In particular, for the strictest cut, $\mu_{\ell^\ast}^\mathrm{cut}=-8.1$~mas~yr$^{-1}$, we recover ${\sim}28\%$ of the bulge population, in good agreement with the ${\sim}30-35\%$ completeness reported by \citet{calamida+2014} and \citet{terry+20} for the equivalent selection in a nearby field.

We stress that these estimates are strongly dependent on the assumed GMM. As discussed in Sect.~\ref{sec:decam_clean}, the model is biased toward the bulge component, so the purity values in Fig.~\ref{Fig:purity_completeness} should be considered as upper limits. More fundamentally, the strong overlap between both components in the VPD implies that kinematics alone cannot fully separate the bulge from the foreground disk. Therefore, any cut in $\mu_{\ell^\ast}$ involves an unavoidable trade-off between purity and completeness, and the optimal choice ultimately depends on the requirements of a given application.

\subsection{The pure bulge color-magnitude diagram}

We used the sample selected with the PM cut $\mu_{\ell^\ast}^\mathrm{cut}=-8.1$~mas~yr$^{-1}$ to analyze the CMD of an almost pure bulge stellar population. After removing the overlapping disk blue plume, the bulge MSTO is visible: Fig.~\ref{Fig:cmd_iso} shows the $i_0,\ (g - i)_0$ CMD of the $\approx$ 10\,000 bulge stars. To compare observations to theory, we used a set of isochrones from the BASTI-IAC database\footnote{\url{http://basti-iac.oa-abruzzo.inaf.it/}} \citep{pietrinferni+2024}.
Medium- to high-resolution spectroscopy from the Giraffe Inner Bulge Survey \citep[GIBS;][]{zoccali+17} shows that most stars in the Baade's window have super-solar metallicities; the distribution shows a main peak at $[{\rm Fe/H}] \approx$ 0.3, a secondary peak at 0 and a more metal-poor tail of stars peaking around $[{\rm Fe/H}] \approx$ -0.4. $[\alpha/\mathrm{Fe}]$ abundances were also measured in the GIBS survey and show that the $\alpha$-enhancement decreases with increasing $[{\rm Fe/H}]$ abundance, reaching $\approx$ 0 at the main metallicity peak of the Baade's window \citep{gonzalez+2015}.
Therefore, we used different sets of models to fit the CMD: a scaled-solar composition is assumed with metallicities of $Z = 0.017$ (solar) and $Z = 0.03$ (super-solar), and an $\alpha$-enhanced one is assumed for the more metal poor models at $Z = 0.012$.
The three panels of Fig.~\ref{Fig:cmd_iso} show the same CMD with over-plotted isochrones for these three metallicities, from top to bottom: $Z = 0.012$, $Z = 0.017$ and $Z = 0.03$, and for different ages, ranging from 2 to 12 Gyrs; Zero Age Horizontal Branch (ZAHB) models for the same metallicities are also over-plotted. 
The photometric catalog used to create the CMD is a subset of the \citet{saha+19} catalog, which was corrected for extinction in all filters by using the color at minimum light of RR Lyrae stars in the field; therefore, only a distance modulus of $DM$ = 14.5 mag was applied to the models. The figure shows that models with solar metallicity best fit the CMD of the average stellar population in the Baade's window; however, to reproduce the full extent of the color and magnitude sequences models at different metallicities are needed.

This figure also shows no evidence of a large fraction ($\approx$ 25\%) of bulge stars younger than 5 Gyr towards the Baade's window. Although a small age spread could be present ($\Delta t \approx$ 2 - 4 Gyr), only a few stars ($\approx$ 3\%) are brighter than the 4 Gyr model. This number is over-estimated due to the MS star sample being not complete. Additionally, some of these stars are BSSs, or residual disk contaminants, further decreasing the fraction of young bulge stars of the CMD.

The clean bulge CMD presented in this study covers a small field of view, $\approx 9\arcmin \times 18\arcmin$, and statistics are still low. However, preliminary results from all other DECam detectors, show similar old MSTO CMDs after applying the PM selection. In order to derive a full detailed SFH of the Baade's window, we will need to run artificial star tests to derive the astrometric and photometric completeness of the dataset. We plan to present this analysis in a forthcoming paper, since this is not the scope of the current one.

\section{Summary} \label{sec:summary}

We presented high-precision PMs toward the Baade's window obtained from DECam observations, combining three epochs (2013, 2021, and 2023) over a ${\sim}$~10-year baseline. As a first step, we restricted our analysis to the field of a single DECam detector, which we adopted as a pilot field to develop and validate a pipeline designed to scale to the full DECam footprint. Our results are summarized as follows.

We developed an astrometric calibration that corrects for the tree-ring pattern and residual field distortion in the resampled images, tying the resulting calibrated positions to the absolute \textit{Gaia}\,DR3 reference frame. The combined tree-ring and polynomial calibration reduced the astrometric residuals from $0.20\arcsec$--$0.40\arcsec$ to $0.02\arcsec$--$0.04\arcsec$, a $>90\%$ improvement in RMS in nearly all observations.

We derived PMs from weighted linear fits to the monthly-averaged positions for stars with at least three epochs and a baseline of at least 8 years. The final catalog contains $48\,239$ stars in the pilot field, with a median PM uncertainty of $0.65$~mas~yr$^{-1}$ for $r_0<18$. Since all positions are tied to \textit{Gaia}\,DR3, the PMs are given in the absolute \textit{Gaia} astrometric system.

We compared our PMs with those from \textit{Gaia}\,DR3 and VIRAC2 in the same region. Our measurements are in excellent agreement with \textit{Gaia}, with slopes close to unity, almost negligible intercepts, and Pearson correlation coefficients above 0.9 across the full magnitude range. Our uncertainties are competitive with, and over most of the magnitude range smaller than, those of VIRAC2. Our catalog also remains nearly complete relative to \citet{saha+19} down to the faintest magnitudes, extending the range over which reliable PMs are available well below the \textit{Gaia} magnitude limit, and roughly one magnitude below the MSTO ($r_0\sim20$).

Using photometrically selected samples, we characterized the kinematics of the Galactic bulge and the foreground disk. For the bulge we find velocity dispersions of $\sigma_{\mu_{\ell^\ast}} = 2.76\pm0.03$~mas~yr$^{-1}$ and $\sigma_{\mu_b} = 2.49\pm0.02$~mas~yr$^{-1}$, with an anisotropy of $1.11\pm0.02$, in agreement with previous HST studies of the Baade's window and nearby fields.

Finally, we used our PMs to obtain a kinematically clean CMD of the bulge reaching below the MSTO. A strict cut in $\mu_{\ell^\ast}$, refined with \textit{Gaia}\,DR3 parallaxes, isolates the bulge population, while a two-component GMM yields a foreground contamination of ${\sim}14\%$ in the field prior to any selection. For the strictest cut, $\mu_{\ell^\ast}<-8.1$~mas~yr$^{-1}$, we recover ${\sim}28\%$ of the bulge with an estimated purity of up to ${\sim}98\%$, consistent with previous HST-based estimates. We stressed that even for the strictest cut, a small fraction (${\sim}2.5\%$) of stars remains in the blue-sequence region of the CMD, likely a combination of residual inner-disk contaminants, bulge blue stragglers, and blue HB stars. Comparison of the cleaned CMD with BaSTI isochrones shows a well-defined old MSTO and no evidence for a large population younger than $\approx$4-5~Gyr in this field. We also caution that the strong overlap between the bulge and disk PM distributions implies that kinematics alone cannot fully separate the two populations, and that complementary information, such as parallaxes or spectroscopic radial velocities, is needed to further reduce the residual contamination.

This pilot study demonstrates that DECam can deliver deep, wide-field PMs reaching below the MSTO from the ground. Extending this analysis to the full DECam footprint will provide kinematically clean bulge samples over much larger areas than currently possible with other instruments, such as HST. Such samples, covering significantly larger regions of the bulge, will enable studies of its star formation history over a wider range of Galactic longitudes and latitudes, rather than being confined to a few small fields. Moreover, since the LSST camera is built on the same thick, fully depleted CCD technology as DECam, it is affected by similar detector-level systematics, most notably the tree-ring pattern. The calibration strategy developed here therefore offers a valuable precursor for crowded-field astrometry with LSST, which will observe the Galactic bulge at unprecedented cadence, opening the possibility of extending PM studies to a large fraction of the bulge.

\begin{acknowledgements}
C.~S.~G. acknowledges support from the National Agency for Research and Development (ANID) Scholarship Program, Doctorado Nacional 2025-21251677. C.~S.~G. also thanks the Space Telescope Science Institute (STScI) for supporting and hosting a visit in 2023, where part of this work was carried out. J.~O.~C. acknowledges financial support from the project ``Partecipazione LSST - Large Synoptic Survey Telescope'' (Ob.Fu. 1.05.03.06; local PI: M. Brescia). Support for M.~C. is provided by ANID’s FONDECYT Regular grant \#1231637 and ANID’s Basal project FB210003. N.~R.~S acknowledges support by the Korea Astronomy and Space Science Institute under the R\&D program (Project No. 2026-1-831-02) supervised by the Korea AeroSpace Administration. We thank M.~L.~Graham for proposing, leading, and carrying out the DECam Deep Drilling Field observations used in this work.
Finally, we also acknowledge the use of the following publicly available software: DAOPHOT \citep{stetson1987}, NumPy \citep{numpy}, SciPy \citep{scipy}, Astropy \citep{astropy}, Matplotlib \citep{matplotlib}, scikit-learn \citep{pedregosa2011} and TOPCAT/STILTS \citep{topcat, stilts}. 
\end{acknowledgements}

\bibliographystyle{aa}
\bibliography{references} 

\renewcommand{\floatpagefraction}{0.9}
\renewcommand{\dblfloatpagefraction}{0.9}

\begin{appendix}

\nolinenumbers
\onecolumn

\section{Tree-rings on the N10 detector}\label{App:N10}

\begin{figure}[ht]
    \centering
    \includegraphics[width=\textwidth]{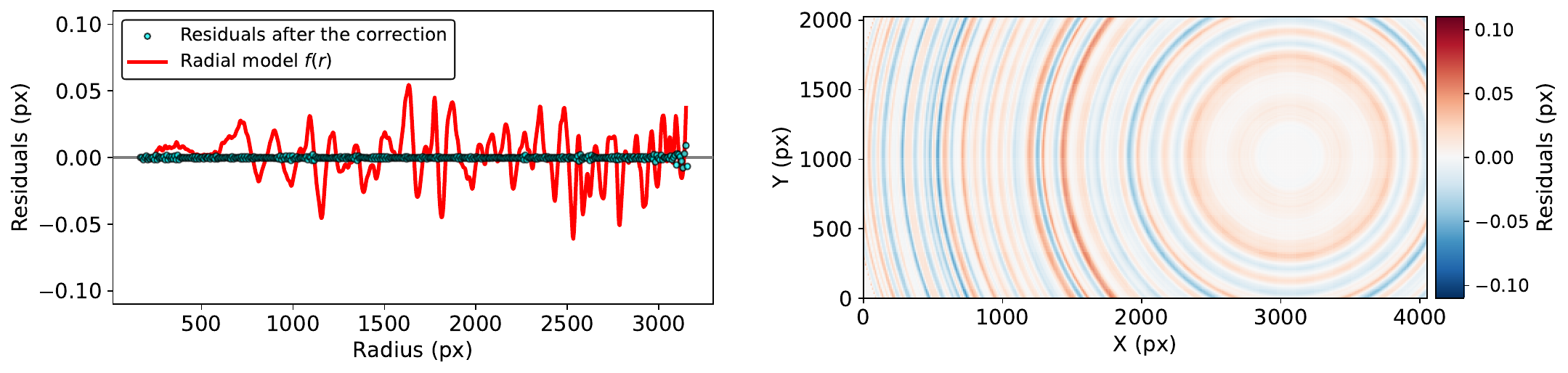}
    \caption{Same as Fig.~\ref{fig:tr_model_s10}, but for the N10 detector, using the combined residuals from individual exposures of} the DECaPS-East 2021 dataset.
    \label{Fig:treerings_N10}
\end{figure}

Following Sect.~\ref{subsec:treerings}, we constructed the tree-ring model for the N10 detector. Since N10 and S10 are different DECam detectors, the tree-ring pattern is modeled independently for each. In contrast to S10, whose ring center lies outside the detector boundary, the center for N10 is located well inside the field, at $(c_X, c_Y) \approx (3067, 1021)$~px, producing a full set of concentric rings within the FoV (see Fig.~\ref{Fig:treerings_N10}).

\section{Uncertainty inflation factor}\label{app:inflation}

\begin{figure}
    \centering
    \includegraphics[width=0.65\textwidth]{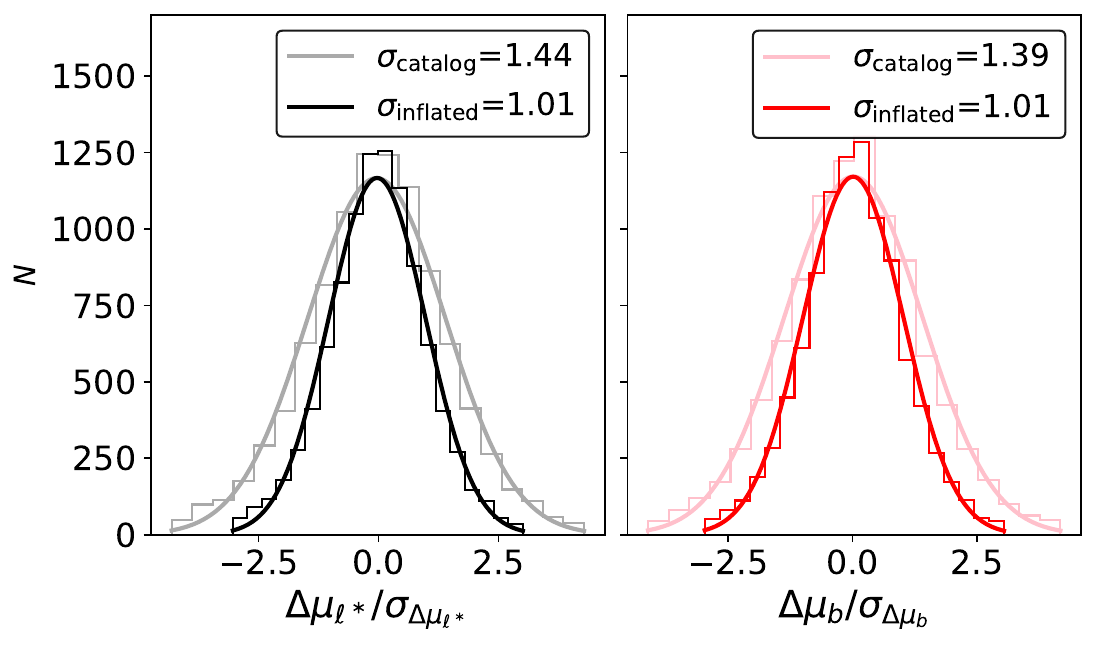}\\[1mm]
    \includegraphics[width=0.70\textwidth]{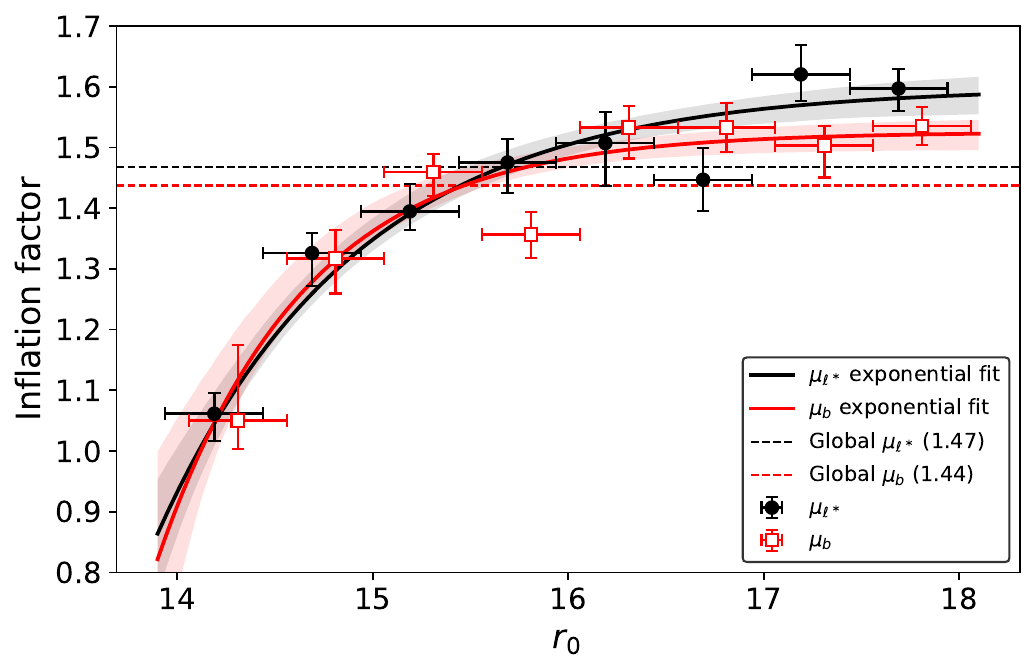}
    \caption{\textit{Top:} Distribution of the PM differences between our DECam catalog and \textit{Gaia}\,DR3, normalized by their combined uncertainties, for $\mu_{\ell^\ast}$ (\textit{left}) and $\mu_b$ (\textit{right}). Grey (pink) histograms show the distributions using the original DECam PM uncertainties, which yield a standard deviation of $\sigma_{\rm catalog}\approx1.4$. Black (red) histograms show the distributions after multiplying the DECam PM uncertainties by the inflation factor (1.47 in $\mu_{\ell^\ast}$ and 1.44 in $\mu_b$), which brings the standard deviation to $\sigma_{\rm inflated}\approx1.0$. The solid curves show the corresponding Gaussian fits, with the standard deviations labeled in each panel. \textit{Bottom:} Inflation factor as a function of $r_0$, computed in bins of $0.5$~mag for $\mu_{\ell^\ast}$ (black circles) and $\mu_b$ (red squares), with $1\sigma$ uncertainties estimated by bootstrap. Solid curves show exponential fits for each component, and dashed lines show the global inflation factors.}
    \label{Fig:inflation_factor}
\end{figure}

\begin{figure}
    \centering
    \includegraphics[width=0.65\textwidth]{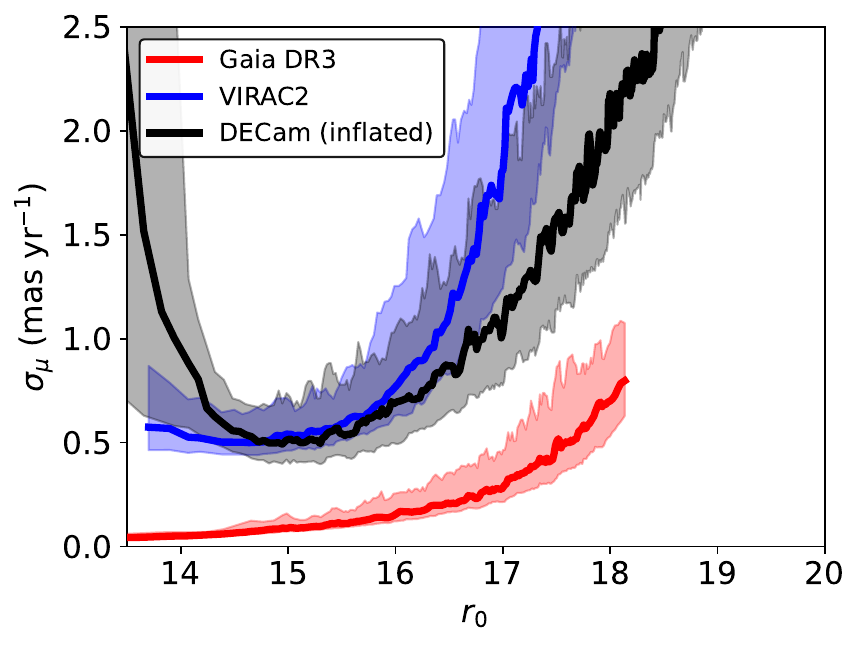}
    \caption{Same as the top panel of Fig.~\ref{Fig:pms_error_comp}, but with the DECam PM uncertainties (black) multiplied by the inflation factor derived in Fig.~\ref{Fig:inflation_factor}.}
    \label{Fig:cse_inflated}
\end{figure}

As detailed in Sect.~\ref{subsec:uncertainties}, our astrometric uncertainties are approximated from the \textsc{DAOPHOT} photometric errors rather than, for example, calibrated with artificial stars. To assess their reliability, we followed the approach of \citet{luna+23} and examined the distribution of the PM differences with respect to \textit{Gaia}\,DR3, normalized by the combined uncertainties of both catalogs. If we assume that the \textit{Gaia}\,DR3 uncertainties are well calibrated, this distribution should follow a unit-width Gaussian, so that any excess width could be attributed to our DECam uncertainties.

As shown in the top panel of Fig.~\ref{Fig:inflation_factor}, the normalized distributions have a standard deviation of $\sigma\approx1.4$ in both coordinates. In principle, this mild excess could originate from either catalog, since the normalization combines the uncertainties of both. However, given the excellent one-to-one agreement between the two catalogs (Sect.~\ref{subsec:gaia_comp}), it more likely suggests that our PM uncertainties are slightly underestimated. Following Eq.~3 from \citet{luna+23}, we derived inflation factors of $1.47$ in $\mu_{\ell^\ast}$ and $1.44$ in $\mu_b$ that bring the standard deviation to ${\sim}1$. Additionally, as proposed by \citet{luna+23}, we also explored the dependence of the inflation factor on magnitude. The bottom panel of Fig.~\ref{Fig:inflation_factor} shows the inflation factor computed in bins of $0.5$~mag over $14 \leq r_0 \leq 18$ with uncertainties estimated by bootstrap. In the brightest bin, the factors are close to unity. However, toward fainter magnitudes, the factors increase and then level off, a behavior that might be described by an asymptotic exponential fit. This is consistent with the increasing impact of crowding toward fainter magnitudes, where unresolved neighbors can perturb the stellar centroids, an effect not captured by our photometric based uncertainties (Sect.~\ref{subsec:uncertainties}). Consequently, the global factors, which are dominated by the more numerous faint stars, overestimate the uncertainties of the brighter stars and slightly underestimate those at $r_0 \gtrsim16.0$. Beyond the \textit{Gaia}\,DR3 magnitude limit, the inflation factors cannot be measured with this method, although the asymptotic behavior of the fit suggests similar values for fainter stars.

Fig.~\ref{Fig:cse_inflated} shows the same PM uncertainty comparison presented in Fig.~\ref{Fig:pms_error_comp}, but after multiplying the uncertainties by the corresponding global inflation factors. Even in this case, our DECam PMs remain competitive with, and lower than, those of VIRAC2 over most of the magnitude range. This suggests that our precision is not solely driven by an underestimation of our uncertainties, and may partly benefit from the finer pixel sampling of DECam. Moreover, adopting instead the maximum inflation factors (see the bottom panel of Fig.~\ref{Fig:inflation_factor}) the conclusions do not change.

\section{Comparison with VIRAC2}\label{app:virac2}

\begin{figure}[ht]
    \centering
    \includegraphics[width=0.75\textwidth]{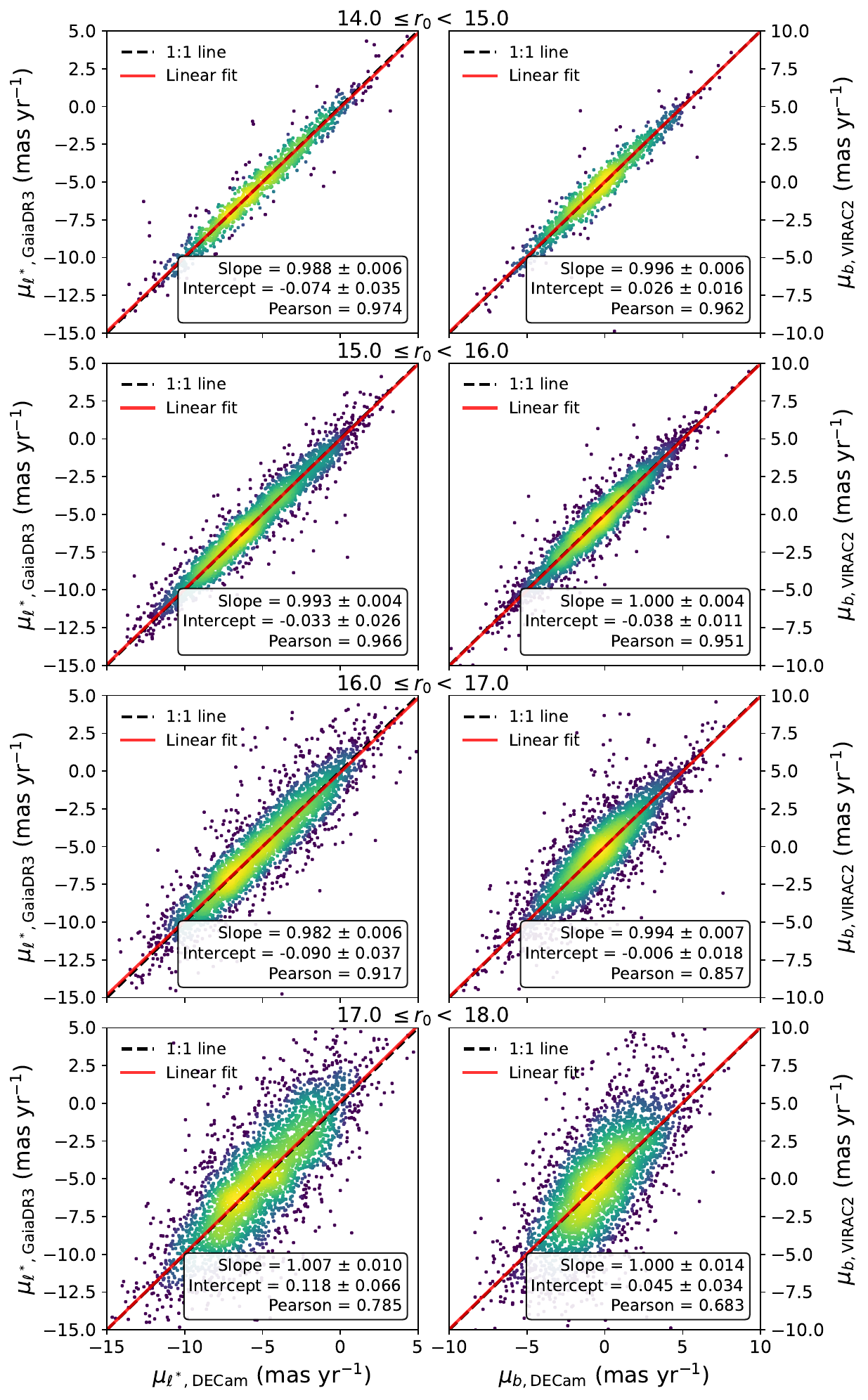}
    \caption{Same as Fig.~\ref{Fig:gaia_comp} but comparing with VIRAC2 instead of \textit{Gaia}\,DR3.}
    \label{Fig:virac2_comp}
\end{figure}

We repeated the same one-to-one comparison described in Sect.~\ref{subsec:gaia_comp}, but comparing our DECam PMs with those from VIRAC2. Since VIRAC2 PMs are also tied to the \textit{Gaia}\,DR3 reference frame, and our astrometric calibration follows the same Chebyshev-polynomial approach adopted by \citet{smith2025} for VIRAC2 (see Sect.~\ref{subsec:chebyshev}), both catalogs are expected to share a common absolute reference frame. This comparison shows good agreement between the two datasets (see Fig.~\ref{Fig:virac2_comp}). For all magnitude bins, the linear fits (red solid lines) show slopes close to unity, ranging from $0.982$ to $1.007$ with no significant trend with magnitude, and almost negligible intercepts, consistent within uncertainties. Pearson correlation coefficients are above 0.85 for $r_0<17.0$, demonstrating strong agreement with VIRAC2. For $r_0>17.0$, they decrease to ${\sim}0.79$ in $\mu_{\ell^\ast}$ and ${\sim}0.68$ in $\mu_b$, as expected given the different nature of both datasets, with VIRAC2 being a NIR survey and DECam an optical instrument, particularly at fainter magnitudes.

\section{Other astrometric systematics}\label{app:systematics}

\begin{figure}
    \centering
    \includegraphics[width=0.65\textwidth]{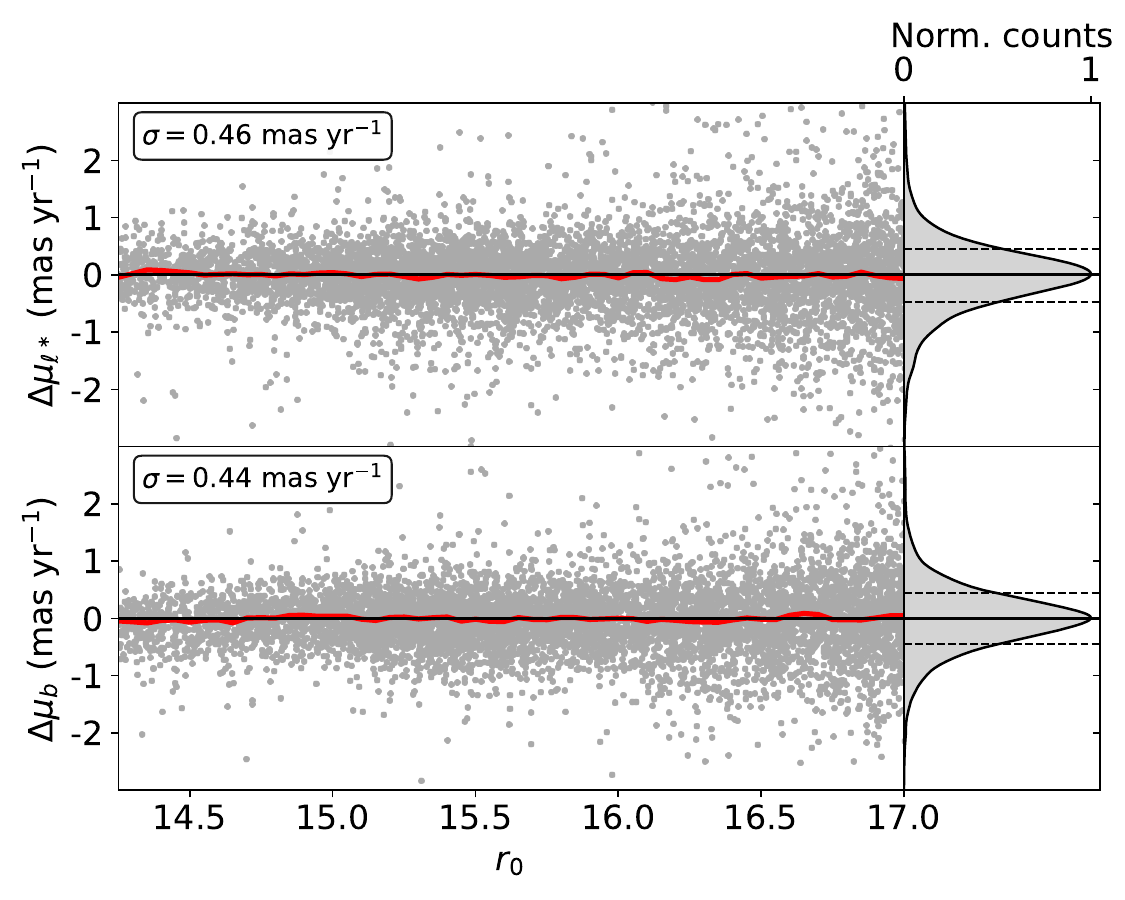}
    \caption{PM differences between our DECam catalog and \textit{Gaia}\,DR3, $\Delta\mu_{\ell^\ast}$ (\textit{top}) and $\Delta\mu_b$ (\textit{bottom}), as a function of $r_0$ magnitude. Grey points show individual stars, and the red line shows the running median. The right panels show the normalized distribution of the differences, with dashed lines marking the $\pm1\sigma$ range, where $\sigma$ is the robust standard deviation, also labeled in the left panels.}
    \label{Fig:deltamu_vs_r}
\end{figure}

\begin{figure}
    \centering
    \includegraphics[width=0.65\textwidth]{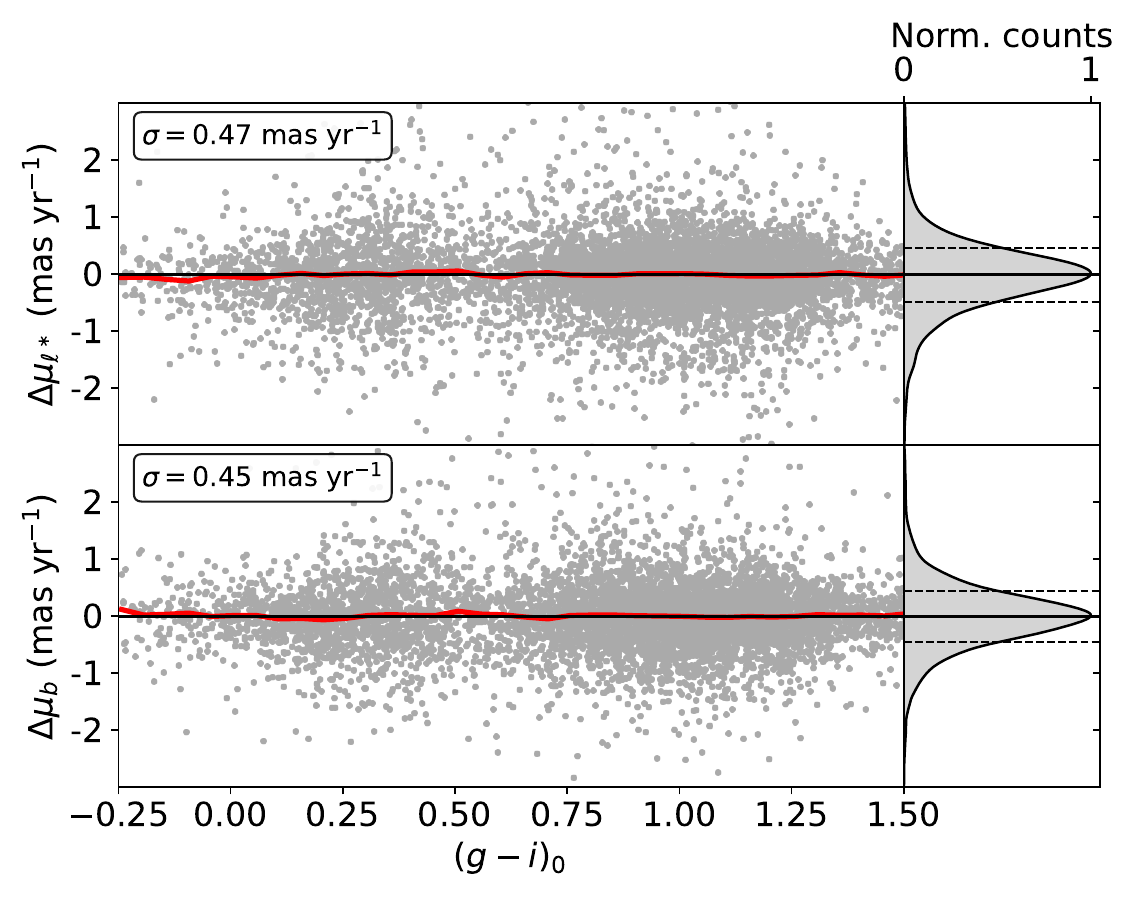}
    \caption{Same as Fig.~\ref{Fig:deltamu_vs_r}, but as a function of the $(g-i)_0$ color.}
    \label{Fig:deltamu_vs_color}
\end{figure}

In this work, we focused primarily on systematics that depend exclusively on the physical position, namely tree-ring pattern, which we modeled with a radial function (Sect.~\ref{subsec:treerings}), and the remaining smooth astrometric distortions (Sect.~\ref{subsec:chebyshev}). However, as shown by \citet{bernstein2017}, DECam astrometry is affected by additional systematics (see their Sects.~4 and 5).

In this appendix, we briefly explore some of the unmodeled systematics through the comparison with \textit{Gaia}\,DR3. First, the residual maps in Fig.~\ref{fig:tr_residuals_s10} reveal an edge distortion along the upper and lower boundaries of the CCD. Second, we also identify some localized increases in the residuals at the corners of the CCD and near the midpoints of each long edge, corresponding to the so-called tape bumps \citep[see Sect.~5 of][]{bernstein2017}. We did not model or correct for these effects, but they do not affect our global astrometry, because they are localized features that can be flagged.

We also examined the dependence of our PM residuals with respect to \textit{Gaia}\,DR3 as a function of magnitude and color. As shown in Fig.~\ref{Fig:deltamu_vs_r}, the residuals remain flat and consistent with zero as a function of $r_0$, ruling out any significant magnitude-dependent systematic. Similarly, Fig.~\ref{Fig:deltamu_vs_color} shows that the residuals are also flat as a function of the $(g-i)_0$ color, indicating that chromatic effects, such as differential chromatic refraction, are not significant. Nevertheless, extending this analysis to the remaining CCDs will require applying all the necessary corrections, given the large variation of the systematics from one detector to another.

\section{Blue-loop candidates}\label{app:blueloop}

\begin{table*}[ht]
    \caption{Astrometric properties and GMM membership probabilities of the blue-loop candidates identified by \citet{saha+19} in our pilot field.}
    \label{Tab:blue_loops}
    \centering
    \small
    \begin{tabular}{r r c c c c c c}
        \hline\hline\noalign{\smallskip}
        RA & Dec & $\mu_{\ell^\ast}$ & $\mu_b$ & $P_{\rm bulge}$ & $P_{\rm disk}$ & $\varpi_{\rm ZP}$ & $d$ \\
        (deg) & (deg) & (mas yr$^{-1}$) & (mas yr$^{-1}$) & & & (mas) & (kpc) \\
        \noalign{\smallskip}\hline\noalign{\smallskip}
        270.703532 & -29.750644 & $-5.96\pm0.57$ & $-0.47\pm0.54$ & 0.90 & 0.10 & $0.28\pm0.03$ & $3.62\pm0.39$ \\
        270.780622 & -29.742568 & $-3.13\pm0.87$ & $ 0.39\pm0.82$ & 0.74 & 0.26 & $0.24\pm0.03$ & $4.12\pm0.54$ \\
        270.790916 & -29.720599 & $ 0.25\pm1.23$ & $-2.77\pm1.18$ & 0.41 & 0.59 & $0.39\pm0.06$ & $2.60\pm0.39$ \\
        270.671841 & -29.791257 & $-1.02\pm1.27$ & $ 3.50\pm1.20$ & 0.78 & 0.22 & $0.23\pm0.03$ & $4.28\pm0.55$ \\
        270.554035 & -29.748079 & $-1.69\pm0.53$ & $-1.32\pm0.51$ & 0.57 & 0.43 & $0.38\pm0.05$ & $2.64\pm0.35$ \\
        270.739751 & -29.744377 & $ 1.63\pm1.87$ & $ 0.00\pm1.78$ & 0.23 & 0.77 & $0.37\pm0.03$ & $2.71\pm0.20$ \\
        270.668044 & -29.716448 & $-11.44\pm0.63$ & $ 0.22\pm0.60$ & 0.99 & 0.01 & $0.28\pm0.03$ & $3.51\pm0.40$ \\
        270.560204 & -29.821476 & $-1.42\pm1.23$ & $ 0.44\pm1.17$ & 0.57 & 0.43 & $0.30\pm0.03$ & $3.28\pm0.32$ \\
        270.870486 & -29.729445 & $-0.98\pm0.43$ & $-3.44\pm0.41$ & 0.59 & 0.41 & $0.26\pm0.03$ & $3.89\pm0.42$ \\
        270.738151 & -29.802788 & $-5.66\pm1.33$ & $-1.66\pm1.25$ & 0.88 & 0.12 & $0.32\pm0.03$ & $3.11\pm0.28$ \\
        270.735803 & -29.761870 & $-0.32\pm2.03$ & $-1.25\pm1.92$ & 0.42 & 0.58 & $0.41\pm0.03$ & $2.47\pm0.17$ \\
        270.570818 & -29.828141 & $-3.62\pm0.74$ & $-0.35\pm0.71$ & 0.76 & 0.24 & $0.34\pm0.03$ & $2.96\pm0.26$ \\
        270.616559 & -29.825008 & $-1.40\pm1.21$ & $ 1.68\pm1.15$ & 0.65 & 0.35 & $0.37\pm0.03$ & $2.69\pm0.21$ \\
        270.642104 & -29.805667 & $ 0.56\pm1.45$ & $ 1.24\pm1.37$ & 0.39 & 0.61 & $0.44\pm0.05$ & $2.29\pm0.27$ \\
        270.560505 & -29.789137 & $ 0.74\pm0.93$ & $ 3.70\pm0.88$ & 0.63 & 0.37 & $0.35\pm0.03$ & $2.87\pm0.24$ \\
        270.747880 & -29.780723 & $ 1.48\pm1.66$ & $ 1.84\pm1.57$ & 0.33 & 0.67 & $0.48\pm0.03$ & $2.07\pm0.12$ \\
        270.822957 & -29.777202 & $-0.25\pm1.57$ & $-3.93\pm1.48$ & 0.56 & 0.44 & $0.34\pm0.03$ & $2.92\pm0.24$ \\
        270.635725 & -29.742340 & $-2.05\pm0.66$ & $ 2.36\pm0.64$ & 0.76 & 0.24 & $0.29\pm0.03$ & $3.45\pm0.36$ \\
        270.614048 & -29.757727 & $-4.47\pm0.39$ & $ 3.17\pm0.38$ & 0.94 & 0.06 & $0.38\pm0.03$ & $2.64\pm0.21$ \\
        270.805202 & -29.785173 & $-5.43\pm2.09$ & $-6.30\pm1.99$ & 0.98 & 0.02 & $0.41\pm0.03$ & $2.45\pm0.15$ \\
        270.571784 & -29.779383 & $-3.71\pm1.30$ & $-7.92\pm1.23$ & 0.99 & 0.01 & $0.36\pm0.03$ & $2.77\pm0.21$ \\
        270.833825 & -29.720844 & $-0.21\pm1.91$ & $-0.84\pm1.83$ & 0.40 & 0.60 & $0.42\pm0.02$ & $2.37\pm0.13$ \\
        270.632276 & -29.811576 & $-0.81\pm0.93$ & $-0.85\pm0.87$ & 0.47 & 0.53 & $0.34\pm0.03$ & $2.91\pm0.26$ \\
        270.655737 & -29.733927 & $-3.37\pm1.42$ & $-0.19\pm1.35$ & 0.74 & 0.26 & $0.39\pm0.03$ & $2.57\pm0.19$ \\
        270.734052 & -29.822848 & $-1.35\pm0.59$ & $-2.09\pm0.55$ & 0.55 & 0.45 & $0.43\pm0.03$ & $2.31\pm0.16$ \\
        270.744314 & -29.828132 & $-3.32\pm1.77$ & $-1.76\pm1.69$ & 0.74 & 0.26 & $0.46\pm0.03$ & $2.18\pm0.13$ \\
        270.564354 & -29.784015 & $-0.26\pm1.28$ & $-3.90\pm1.17$ & 0.56 & 0.44 & $0.32\pm0.03$ & $3.09\pm0.28$ \\
        270.572659 & -29.772039 & $-7.62\pm2.12$ & $ 3.81\pm1.99$ & 0.99 & 0.01 & $0.23\pm0.03$ & $4.38\pm0.62$ \\
        270.556742 & -29.761296 & $-2.39\pm1.55$ & $-2.52\pm1.48$ & 0.68 & 0.32 & $0.31\pm0.03$ & $3.24\pm0.27$ \\
        270.727915 & -29.839853 & $-10.92\pm2.02$ & $-1.02\pm1.91$ & 0.98 & 0.02 & $0.52\pm0.03$ & $1.94\pm0.10$ \\
        270.613234 & -29.752032 & $-5.92\pm0.36$ & $ 0.93\pm0.35$ & 0.92 & 0.08 & $0.29\pm0.03$ & $3.41\pm0.38$ \\
        \hline
    \end{tabular}
    \tablefoot{RA and Dec are given in degrees, while PMs and their uncertainties are expressed in Galactic coordinates. The membership probabilities $P_{\rm bulge}$ and $P_{\rm disk}$ were derived from the two-component GMM using PMs only. The Gaia DR3 parallaxes ($\varpi_{\rm ZP}$) were corrected for the zero-point offset. Distances were estimated by direct parallax inversion, $d\,(\mathrm{kpc}) = 1/\varpi_{\rm ZP}\,(\mathrm{mas})$, with uncertainties obtained through standard error propagation.}
\end{table*}

\end{appendix}

\end{document}